\documentclass[
 reprint,
nofootinbib,
 amsmath,amssymb,
 aps,
]{revtex4-2}

\usepackage{graphicx, color}
\usepackage{dcolumn}
\usepackage{bm}
\usepackage[mathlines]{lineno}
\usepackage{amsmath}
\usepackage{amssymb}
\usepackage{float}
\usepackage{mathrsfs}
\usepackage{lipsum}
\usepackage{hyperref}
\usepackage[normalem]{ulem}
\hypersetup{
    colorlinks=true,       
    linkcolor=red,          
    citecolor=rp,        
    filecolor=magenta,      
    urlcolor=blue           
}
\usepackage{cleveref}
\usepackage{color}

\newcommand{\mpl}{m_\mathrm{pl}}

\definecolor{rp}{cmyk}{0.2, 1, 0.6, 0}
\definecolor{rp}{cmyk}{0.2, 1, 0.6, 0}
\definecolor{green2}{cmyk}{0.27, 0, 1, 0.52}

\newcommand{\bPsi}{{\bm{\mathsf{\Psi}}}}
\newcommand{\bepsilon}{{\bm{\mathsf{\epsilon}}}}
\newcommand{\bS}{{\bm{S}}}
\newcommand{\bW}{{\bm{W}}}
\newcommand{\bH}{{\bm{\mathsf{H}}}}
\newcommand{\bR}{{\bm{\mathsf{R}}}}
\DeclareMathOperator{\Tr}{Tr}

\usepackage{hyperref}
\hypersetup{%
  colorlinks = true,
  linkcolor  = rp
}

\begin{document}

\preprint{APS/123-QED}

\title{Polarized solitons in higher-spin wave dark matter}

\author{Mudit Jain}
\email{mudit.jain@rice.edu}
\author{Mustafa A. Amin}%
 \email{mustafa.a.amin@rice.edu}
\affiliation{%
Department of Physics and Astronomy, Rice University\\
Houston, TX, 77005, U.S.A.
}%

\date{\today}

\begin{abstract}
We first show that the effective non-relativistic theory of gravitationally interacting, massive integer-spin fields (spin-$0$, $1$, and $2$ in particular) is described by a $2s+1$ component Schr\"{o}dinger-Poisson action, where $s$ is the spin of the field. We then construct $s+1$ distinct, gravitationally supported solitons in this non-relativistic theory from identically polarized plane waves. Such solitons are extremally polarized, with macroscopically large spin, but no orbital angular momentum. These $s+1$ solitons form a basis set, out of which partially polarized solitons can be constructed. All such solitons are ground states, have a spherically symmetric energy density but not field configurations.  We discuss how solitons in higher-spin fields can be distinguished from scalar solitons, and potential gravitational and non-gravitational probes of them.
\end{abstract}

\maketitle

\section{Introduction}
Dark matter makes up approximately $84\%$ of the non-relativistic matter in our cosmos \cite{Planck:2018vyg}, but its detailed nature remains uncertain. For example, the mass of the fundamental particles making up dark matter can range from $\sim 10^{-21}\rm eV$ \cite{Irsic:2017yje} to  $\sim \mpl$. The upper bound is extended further if dark matter is multi-component, or composite \cite{Jacobs:2014yca}. We also have no robust constraints on the spin of the particles/fields that make up dark matter.  What we do know from a plethora of observations is that dark matter is very weakly interacting with the Standard model, is non-relativistic in the contemporary universe, and has clumped efficiently under the influence of gravity (for a historical overview, see \cite{Bertone:2016nfn}). \\

With the aim of better understanding dark matter, we wish to explore how the fundamental degrees of freedom, such as mass and spin of the dark field can manifest themselves on macroscopic scales. As an extreme case for the mass, making the dark matter particle ultra-light is one way to boost its Compton/de Broglie scale to  astrophysical scales, thus allowing us to probe astrophysical effects sensitive to this mass \cite{Niemeyer:2019aqm,Grin:2019mub,Ferreira:2020fam}.\\

\begin{figure*}[t]
\begin{center}
\includegraphics[width=1\textwidth]{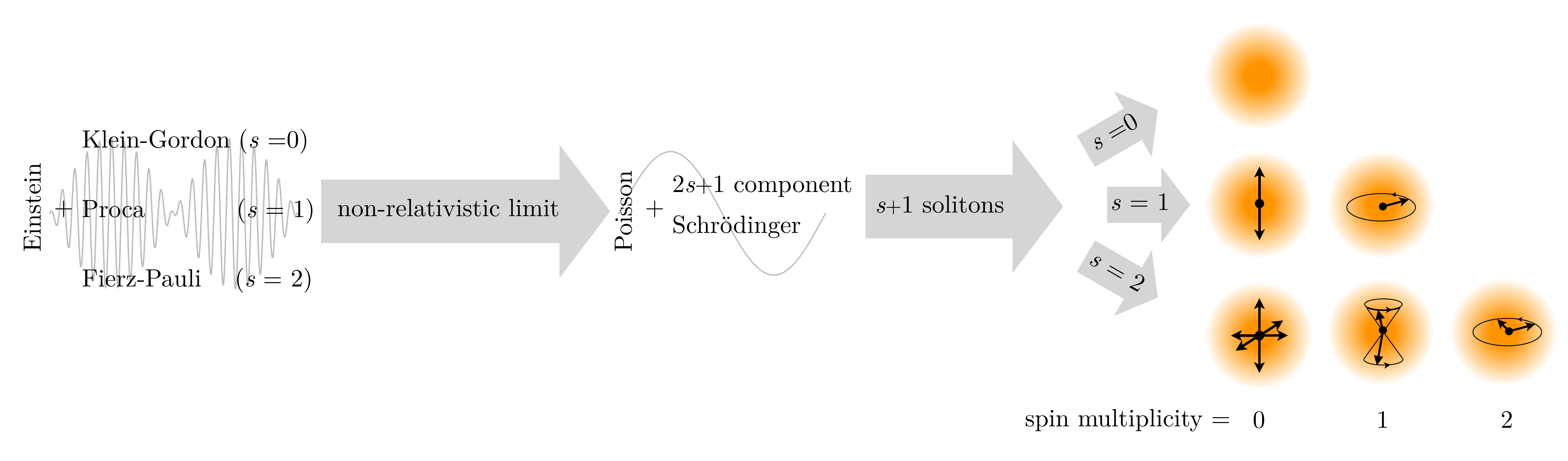}
\end{center}
\caption{A visual summary of some of the main results of our paper.} 
\label{fig:schematic}
\end{figure*}

In this paper, we are interested in whether the spin of the dark fields can also have implications on macroscopic scales. Since we are interested in this field being all or part of dark matter, we first derive the non-relativistic limit of massive, higher-spin bosonic field that is capable of clustering under gravity. We show that the non-relativistic behaviour of a spin-$s$ bosonic field is described by a $2s+1$ component Schr\"{o}dinger-Poisson system. 
Our results are also applicable to fields in the early universe, for example, at the end of inflation \cite{Amin:2014eta,Lozanov:2019jxc} or later \cite{Allahverdi:2020bys}.

Fields with non-linear interactions (gravitational or self-interactions) often allow for coherent, metastable field configurations called solitons. With this non-relativistic limit of our higher-spin dark fields at hand, we derive the lowest energy soliton solutions in spin-$0$, spin-$1$ and spin-$2$ fields. In general for a spin-$s$ field, we have $s+1$ distinct and ``extremally" polarized solitons with equal energy, where all the waves making up a given soliton have the same multiplicity of spin (which we refer to as polarization). We show that the total intrinsic spin of such solitons can be macroscopically (even astrophysically) large, while still having zero orbital angular momentum and a spherical energy density. We also construct fractionally polarized solitons from these extremally polarized ones. All such solitons have a universal mass density profile (up to scalings). We call all such ground-state solitons ``$p$-solitons". For a visual summary of the main results of our paper, see Fig.~\ref{fig:schematic}. We also discuss higher energy solitons, and compare them to our $p$-solitons.\\

The non-relativistic limit of scalar (spin-$0$) dark matter has of course been explored before. For recent works, including relativistic corrections, see \cite{Namjoo:2017nia,Salehian:2020bon,Salehian:2021khb}. The non-relativistic limit of vector (spin-$1$)  dark matter in an expanding universe was discussed in \cite{Adshead:2021kvl}, whereas the tensor (spin-$2$) case in an asymptotically flat spacetime was discussed in \cite{Aoki:2017ixz}. In our work, we recover these results and provide a  framework that allows for a generalization to higher spins.\footnote{The massive spin-$2$ case comes with subtleties related to having two metrics, self-interactions, limiting the coupling to the Standard model for these fields to be dark, as well as having a consistent theory at sufficiently high energies. Some of these are discussed in the Appendix.}\\

Similarly, different types of solitons in the non-relativistic limit of massive spin-$0$ \cite{PhysRev.187.1767,Chavanis_2011}, spin-$1$ \cite{Adshead:2021kvl} and spin-$2$ field \cite{Aoki:2017ixz} appear in the literature. 
A distinguishing feature of our analysis is that we construct solitons with a focus on the polarization of their constituent plane waves. This lead us to uncover a novel class of extremally polarized solitons with spin $S_{\rm tot}/\hbar=\lambda M/m$ which can be macroscopically large for $M\gg m$.  Here, $m$ is the mass of the field, $M$ is the mass of the soliton and $\lambda$ is the spin multiplicity. These coherent solitons (along with fractionally polarized ones mentioned earlier) might open up new avenues for observationally probing higher-spin fields.\\

We find that even within Newtonian gravity it might be possible to distinguish interacting solitons with different polarizations. Going beyond Newtonian gravity, which we do not pursue here, might remove degeneracies between different polarizations of the higher-spin fields even further. We also discuss possibilities of probing higher-spin dark matter via non-gravitational interactions astrophysically, taking advantage of the polarization state of the solitons. In this case, various terrestial experiments~\cite{Caldwell:2016dcw,Baryakhtar:2018doz,Chiles:2021gxk} may be used for detectable signatures.\\ 

The paper is organized as follows. In section~\ref{sec:models} we discuss our model for the case of dark scalar, vector, and tensor massive fields, leaving additional details in Appendix~\ref{sec:Appendix}. In section~\ref{sec:NR_effective} we provide the effective non-relativistic action (which is the Schr\"{o}dinger-Poisson system) for these dark integer spin fields, and discuss the various symmetries of the action. In section~\ref{sec:grav_bound} we discuss the  gravitationally bound solitons. In section~\ref{sec:distinguish} we discuss their distinguishability, primarily within Newtonian gravity, and also mention other non-gravitational couplings that can probe the spin nature of the fields. In section~\ref{sec:sum_fd} we summarize and also highlight some future directions worth investigating.

\section{Models}\label{sec:models}
Our matter Lagrangian consists of the usual Standard Model (SM) sector, along with some dark sector that includes additional massive spin-$0$, spin-$1$, or even spin-$2$ fields. We take these fields to be real valued. \\

Explicitly, our general action has the form 
\begin{align}
    S = S_{\rm EH} + S_{\rm dark} + S_{\rm vis}\,,
\end{align}
where $S_{\rm EH}$ is the gravity sector, $S_{\rm dark}$ is some dark sector (incluing dark integer spin fields), and $S_{\rm vis}$ is the visible sector (comprising of the SM). Our focus is only on the gravity + dark sector in this paper. We consider perturbations of different fields around some background metric $\bar{g}_{\mu\nu}$ which leads to the usual massless spin-$2$ fluctuations: $h_{\mu\nu}$ (the graviton), along with other perturbations in different fields. We will focus on a given spin-$s$ field + gravity, instead of considering massive spin-$0$, $1$ and $2$ together, although our formalism can accomodate the latter scenario as well.\\

For most part, we are interested in sub-horizon physics where length scales associated with configurations of these dark fields are much smaller than the Hubble horizon.  As a result, we ignore Hubble expansion, and take the background metric to be\footnote{We use $\bar{g}_{\mu\nu}  = \mathrm{diag}(1,-a^2(t),-a^2(t),-a^2(t))$ for an expanding universe when needed. Here, $a(t)$ is the scale factor normalized to unity today.} $\bar{g}_{\mu\nu} = \eta_{\mu\nu} = \mathrm{diag}(1,-1,-1,-1)$. We also take $\hbar=c=1$.\\

In the next three subsections, we provide the general action up-to quadratic order in the fields of interest, along with leading order gravitational interactions. For the non-relativistic limit that we are interested in, the leading order actions provided here are sufficient. The full nonlinear actions are discussed in the Appendix. 
\subsection{Spin-$0$}
The quadratic (free) action for the spin-$0$ field $\phi$, and metric fluctuations $h_{\mu\nu}$,  along with their leading interactions is given by
\begin{align}\label{eq:0-actions2}
    S_{\rm EH} + S_{\rm dark} &= \int\mathrm{d}^4x\Bigl[\mpl^2\mathcal{L}^{(2)}_{\rm GR}(h) + \mathcal{L}^{(2)}_{m,0}(\phi)\nonumber\\
    &\qquad\qquad\quad - \frac{1}{2}h_{\mu\nu}\mathcal{T}^{\mu\nu}(\phi) + ...\Bigr],
\end{align}
where $\mathcal{L}^{(2)}_{\rm GR}(h)$ is the linearized Einstein-Hilbert action:
\begin{align}\label{eq:linearized_EH}
    \mathcal{L}^{(2)}_{\rm GR}(h) &= \frac{1}{4}\Bigl\{\eta^{\alpha\beta}\eta^{\lambda\sigma}\eta^{\kappa\rho}\partial_{\beta}h_{\sigma\rho}\partial_{\alpha}h_{\lambda\kappa}  - \eta^{\alpha\beta}\partial_{\beta}h\partial_{\alpha}h\nonumber\\
    & + 2\eta^{\lambda\beta}\eta^{\kappa\alpha}\partial_{\beta}h\partial_{\alpha}h_{\lambda\kappa} - 2\eta^{\kappa\rho}\eta^{\sigma\beta}\eta^{\lambda\alpha}\partial_{\beta}h_{\sigma\rho}\partial_{\alpha}h_{\lambda\kappa}\Bigr\},
\end{align}
and $\mathcal{L}^{(2)}_{m,0}(\phi)$ and $\mathcal{T}_{\mu\nu}(\phi)$ are the usual Klein-Gordon Lagrangian density and leading order energy-momentum tensor:
\begin{align}\label{eq:KG_system}
    \mathcal{L}^{(2)}_{m,0}(\phi) &= \frac{1}{2}\eta^{\mu\nu}\partial_{\mu}\phi\,\partial_{\mu}\phi - \frac{1}{2}m^2\phi^2\,,\nonumber\\
    \mathcal{T}_{\mu\nu}(\phi) &= \partial_{\mu}\phi\,\partial_{\nu}\phi - \eta_{\mu\nu}\,\mathcal{L}^{(2)}_{m,0}(\phi).
\end{align}
The `$...$' in \eqref{eq:0-actions2} represents higher-order interaction terms between $\phi$ and $h$ along with other dark sector fields. Self-interactions can easily be included by adding $V_{\rm nl}(\phi)=\lambda_3 \phi^3+\lambda_4 \phi^4\hdots$ to the quadratic potential , but we ignore them in this paper.\\ 

While the massive spin-$0$ field has no constraints, the gravitational sector does. We discuss these constraints and the action for the physical degrees of freedom in the gravitational sector at the end of this  section. 
\subsection{Spin-$1$}
For a massive spin-$1$ field $W_\mu$, we have the following leading order action for $h_{\mu\nu}$ and $W_{\mu}$:
\begin{align}\label{eq:L_spin1}
    S_{\rm EH} + S_{\rm dark} &= \int\mathrm{d}^4x\Bigl[\mpl^2\mathcal{L}^{(2)}_{GR}(h) + \mathcal{L}^{(2)}_{m,1}(W)\nonumber\\
    &\qquad\qquad\quad - \frac{1}{2}h_{\mu\nu}T^{\mu\nu}(W) + ...\Bigr],
\end{align}
where $G_{\mu\nu}=\partial_\mu W_\nu-\partial_\nu W_\mu$; $\mathcal{L}^{(2)}_{m,1}(W)$ and $T_{\mu\nu}(W)$ are the usual Proca Lagrangian density and leading order energy momentum tensor:
\begin{align}\label{eq:Proca_system}
    \mathcal{L}^{(2)}_{m,1}(W) &= -\frac{1}{4}\eta^{\mu\alpha}\eta^{\nu\beta}\,G_{\mu\nu}G_{\alpha\beta} + \frac{1}{2}m^2\,\eta^{\mu\nu}W_{\mu}W_{\nu}\nonumber\\
    T_{\mu\nu}(W) &= -\eta^{\alpha\beta}G_{\mu\alpha}G_{\nu\beta} + m^2W_{\mu}W_{\nu} - \eta_{\mu\nu}\,\mathcal{L}^{(2)}_{m,1}(W).
\end{align}
Similar to the case of the scalar field, we can include self-interactions via $V_{\textrm{nl}}=-\lambda_4 (W_\mu W^\mu)^2+\lambda_6(W_\mu W^\mu)^3+\hdots$.  As we shall show elsewhere~\cite{Zhang:2021xxa}, such a setup with $\lambda_4,\lambda_6>0$ can arise as an effective theory, and admits vector oscillon solutions supported by self-interactions. However, for our present purposes, we assume no self-interactions.\\

\subsubsection*{Action for the physical d.o.f}
To aid the transition to the non-relativistic limit, we eliminate the constraints in the massive spin-$1$ sector and provide the action for the three physical degrees of freedom (namely the three spin-multiplicity states). 
Varying the action~\eqref{eq:L_spin1} with respect to $W_{0}$ and substituting back, yields\footnote{This is a straightforward, but long exercise which we omit here to reduce clutter.}
\begin{align}\label{eq:physical_Ldens_s1}
    &\mathcal{L}^{(2)}_{m,1} - \frac{1}{2}h_{\mu\nu}T^{\mu\nu}(W)\nonumber\\
    &=\frac{1}{2}\dot{W}_i\,\mathcal{P}_{ij}(m)\,\dot{W}_j - \frac{1}{2}W_i\left(-\nabla^2 + m^2\right)\,\mathcal{P}_{ij}(m)\,W_j\nonumber\\
    &\quad -\frac{1}{2}h_{\mu\nu}\mathcal{T}^{\mu\nu},
\end{align}
where $\mathcal{T}^{\mu\nu}$ is the energy momentum tensor (with $W_0$ substituted in \eqref{eq:Proca_system}), and $\mathcal{P}_{ij}(m)$ is the projection operator 
\begin{align}\label{eq:projection_operator}
    \mathcal{P}_{ij}(m) \equiv \delta_{ij} + \frac{\partial_i\partial_j}{-\nabla^2 + m^2}\,.
\end{align}

As mentioned earlier, the gravitational sector constraints are discussed at the end of this section.
\subsection{Spin-$2$}\label{sec:spin-2_models}
In the case of a massive spin-$2$ field $H_{\mu\nu}$, we take the quadratic actions for $h_{\mu\nu}$ and $H_{\mu\nu}$ + leading interactions to be:
\begin{align}\label{eq:bi-gravity_expanded}
    S_{h,H} &= \int\mathrm{d}^4x\Bigl\{\mpl^2\mathcal{L}^{(2)}_{GR}(h) + \mathcal{L}^{(2)}_{m,2}(H)\nonumber\\ 
    &\qquad\qquad\qquad - \frac{1}{2}h_{\mu\nu}T^{\mu\nu}(H) + ...\Bigr\},
\end{align}
where $\mathcal{L}^{(2)}_{m,2}(H)$ is the Fierz-Pauli Lagrangian density and $T^{\mu\nu}(H)$ is the associated energy momentum tensor:
\begin{align}\label{eq:FP_system}
    \mathcal{L}^{(2)}_{m,2}(H) &= \frac{1}{2}\eta^{\alpha\beta}\eta^{\lambda\sigma}\eta^{\kappa\rho}\partial_{\beta}H_{\sigma\rho}\partial_{\alpha}H_{\lambda\kappa} - \frac{1}{2}\eta^{\alpha\beta}\partial_{\beta}H\partial_{\alpha}H\nonumber\\
    &- \eta^{\kappa\rho}\eta^{\sigma\beta}\eta^{\lambda\alpha}\partial_{\beta}H_{\sigma\rho}\partial_{\alpha}H_{\lambda\kappa} + \eta^{\lambda\beta}\eta^{\kappa\alpha}\partial_{\beta}H\partial_{\alpha}H_{\lambda\kappa}\nonumber\\
    &+ \frac{1}{2}m^2\left[\eta^{\mu\nu}\eta^{\alpha\beta}H_{\mu\nu}H_{\alpha\beta} - \eta^{\sigma\lambda}\eta^{\rho\kappa}H_{\sigma\rho}H_{\lambda\kappa}\right];\nonumber\\
    \nonumber\\
    T_{\mu\nu}(H) &= 
    \Bigl\{\partial_\mu H_{\sigma\rho}\partial_\nu H^{\sigma\rho} + 2\eta^{\kappa\rho}\partial^\alpha H_{\nu\rho}\partial_\alpha H_{\mu\kappa}\nonumber\\& - 2\partial^\beta H_{\beta\nu}\partial^\alpha H_{\alpha\mu} - 2\partial_\nu H_{\mu\rho}\partial_\alpha H^{\alpha\rho} + \partial_\nu H \partial^\alpha H_{\mu\alpha}\nonumber\\
    &+ \partial^\alpha H \partial_\nu H_{\alpha\mu} - 2\partial_\mu H_{\nu\rho}\partial_\alpha H^{\alpha\rho} +\partial_\mu H \partial^\alpha H_{\nu\alpha}\nonumber\\
    &+ \partial^\alpha H \partial_\mu H_{\alpha\nu} - \partial_\mu H\partial_\nu H\Bigr\}\nonumber\\
    &- 2m^2\left(\eta^{\alpha\beta} H_{\mu\alpha}H_{\nu\beta} - H_{\mu\nu}H\right)\nonumber\\
    & -2\partial_\xi\Bigl\{H_{\kappa\mu}\partial_\nu H^{\kappa\xi} + H_{\kappa\nu}\partial_\mu H^{\kappa\xi} + \eta^{\kappa\rho}H_{\kappa\mu}\partial^{\xi}H_{\nu\rho}\nonumber\\
    &+ \eta^{\kappa\rho}H_{\kappa\nu}\partial^{\xi}H_{\mu\rho} - H^{\alpha\xi}\partial_\nu H_{\mu\alpha} - H^{\alpha\xi}\partial_\mu H_{\nu\alpha}\nonumber\\
    &- 2\delta^{\xi}_{\nu}H_{\kappa\mu}\partial_\beta H^{\beta\kappa} - 2\delta^{\xi}_{\mu}H_{\kappa\nu}\partial_\beta H^{\beta\kappa} - H_{\mu\nu}\partial_\beta H^{\beta\xi}\nonumber\\
    &+ H_{\mu\nu}\partial^\xi H + \delta^{\xi}_{\nu}H_{\alpha\mu}\partial^\alpha H + \delta^{\xi}_{\mu}H_{\alpha\nu}\partial^\alpha H\nonumber\\
    &- \eta_{\mu\nu}H^{\alpha\xi}\partial_\alpha H\Bigr\} - \eta_{\mu\nu}\,\mathcal{L}^{(2)}_{m,2}(H)\,.
\end{align}
At this leading order, we obtained the energy momentum tensor by promoting $\eta_{\mu\nu}$ to $g_{\mu\nu} = \eta_{\mu\nu} + h_{\mu\nu}$ in $\mathcal{L}^{(2)}_{m,2}(H)$, along with covariant derivatives with respect to $g$, and then simply reading out the terms that couple to $h_{\mu\nu}$. That is, simply minimally coupling it to gravity just like any other matter field. We also arrive at the above action from the bi-gravity theory \cite{deRham:2010kj,Hassan:2011hr,Hassan:2011tf,Hassan:2011zd,Hinterbichler:2011tt,deRham:2014zqa,Schmidt-May:2015vnx} (which is a ghost-free, nonlinear completion of our quadratic action) in the Appendix.\\

The massive spin-$2$ field scenario includes additional subtleties compared to the lower spin cases. Unlike the spin-$0$ and spin-$1$ case, self-interactions are unavoidable in the massive spin-$2$ sector within bi-gravity. However,  they are conveniently suppressed in the non-relativistic limit (see discussion around eq. \eqref{eq:self_int_vertices} in the Appendix). Furthermore, for the massive spin-$2$ field to be dark, its direct couplings to the visible sector that arise in the bi-gravity theory need to be suppressed. This is also possible using a particular choice of an effective metric \cite{Noller:2014sta,Bonifacio:2017nnt} that couples to the SM. We discuss this further in Sec.~[\ref{sec:coup_matter}] of the Appendix. Within the bi-gravity context, the extension to FLRW background is also non-trivial, and might have instabilities in the early universe \cite{Comelli:2015pua} when using the effective metric mentioned above. For our purposes, we remain agnostic regarding a full non-linear interacting theory of a massive spin-$2$ dark field in an FLRW background, and simply work with the above quadratic action.
\subsubsection*{Actions for physical d.o.f}
For massive tensor field $H_{\mu\nu}$, the constrained variables are $H_{00}$, $H_{0i}$ and $\Tr[H_{ij}]$, leaving 5 physical degrees of freedom (the 5 spin-multiplicity states). After solving for these constraints (obtained by varying the action~\eqref{eq:bi-gravity_expanded} with respect to these variables), and plugging them back into the action, we get the following Lagrangian density
\begin{align}\label{eq:physical_Ldens_s2}
    &\mathcal{L}^{(2)}_{m,2} - \frac{1}{2}h_{\mu\nu}T^{\mu\nu}\nonumber\\
    &= \frac{1}{2}\dot{H}_{ij}\,\mathcal{P}_{ik}(m)\,\mathcal{P}_{jl}(m)\,\dot{H}_{kl}\nonumber\\
    &- \frac{1}{2}H_{ij}\,\mathcal{P}_{ik}(m)\,(-\nabla^2+m^2)\,\mathcal{P}_{jl}(m)\,H_{kl}\nonumber\\
    & -\frac{1}{2}h_{\mu\nu}\mathcal{T}^{\mu\nu}.
\end{align}
Here, $\mathcal{P}_{ij}$ is the projection matrix~\eqref{eq:projection_operator}, and $\mathcal{T}_{\mu\nu}(H_{ij})$ is the energy momentum tensor (dependent on the physical degrees of freedom only).\\

\subsection{The gravitational sector}
Recall that  the action has the general form:
\begin{align}\label{eq:0-actions-s}
     S_{\rm EH} + S_{\rm dark} &= \int\mathrm{d}^4x\Bigl[\mpl^2\mathcal{L}^{(2)}_{\rm GR}(h) + \mathcal{L}^{(2)}_{m,s}(\bm{\mathcal{F}})\nonumber\\
     &\qquad\qquad\quad - \frac{1}{2}h_{\mu\nu}\mathcal{T}^{\mu\nu}(\bm{\mathcal{F}}) + ...\Bigr],
 \end{align}
 where $\bm{\mathcal{F}}$ now represents the physical degrees of freedom in the spin-$s$ field, and $\mathcal{T}^{\mu\nu}$ is the corresponding energy-momentum tensor. Varying this action with respect to $h_{0i}$, $\mathrm{Tr}[h_{ij}]$, and $h_{00} \equiv 2\Phi$, we obtain
\begin{align}\label{eq:h_constraints}
    \nabla^2\,\mathcal{P}_{ij}(0)\,h_{0i} &= \partial_i\dot{h}_{ij}  -\partial_j{\rm Tr}[\dot{h}_{ij}] - \frac{\mathcal{T}_{0j}}{\mpl^2}\,,\nonumber\\
    \nabla^2{\rm Tr}[h_{ij}] &= \partial_i\partial_jh_{ij} +  \frac{\mathcal{T}_{00}}{\mpl^2}\,,\nonumber\\
    2\,\nabla^2\Phi &= -\frac{\partial_i\partial_j}{\nabla^2}\,\ddot{h}_{ij} - \Box \frac{1}{\nabla^2} \frac{\mathcal{T}_{00}}{\mpl^2}\,.
\end{align}
When these constraints are plugged back in the above action, we get the quadratic action for the physical degrees of freedom in the gravitational sector + interactions. The  quadratic action is the same as the first two terms on the right hand side of \eqref{eq:physical_Ldens_s2}, with  $m=0$ and $H_{ij}\rightarrow h_{ij}$\footnote{The projection operator $P_{ij}(0)$, along with the constraints, only allows for the 2 physical degrees of freedom in $h_{ij}$ (which are the transverse and traceless gravitational waves).}, and the interaction term is
\begin{align}\label{eq:interaction_physical}
    \frac{1}{2}\,h_{\mu\nu}\,\mathcal{T}^{\mu\nu} &= \frac{1}{2}h_{ij}\Biggl\{\mathcal{T}_{ij} + \partial_i\partial_j\frac{1}{\nabla^2}\frac{1}{\nabla^2}\ddot{\mathcal{T}}_{00} - \partial_j\frac{1}{\nabla^2}\dot{\mathcal{T}}_{0i}\nonumber\\
    &\qquad\quad - \partial_i\frac{1}{\nabla^2}\dot{\mathcal{T}}_{0j} + 2\,\partial_i\partial_j\frac{1}{\nabla^2}\partial_k\frac{1}{\nabla^2}\dot{\mathcal{T}}_{0k}\Biggr\}.
\end{align}
\section{Effective non-relativistic theory}\label{sec:NR_effective}
With the Lagrangian densities at hand from the previous section, we now decompose $\bm{\mathcal{F}}$ (which represents the physical d.o.f of different integer spin-fields and  carries one and two spatial indices for spin-$1$ and spin-$2$ cases respectively) in the following fashion
\begin{align}\label{eq:F_decompose}
    \bm{\mathcal{F}}({\bf x},t) = \frac{1}{\sqrt{2m}}\left[e^{-imt}\tilde{\bPsi}({\bf x},t) + \mathrm{h.c.}\right].
\end{align}
To obtain the non-relativistic limit, we work with the slowly varying piece in $\tilde{\bPsi}$ that we denote as $\bPsi$. Essentially, we discard all the terms that carry the oscillating factor $e^{\pm i2mt}$ and two time derivative terms in the action (which would be suppressed by factors of $k/m$, where $k$ is a characteristic wave-number). The projection operator \eqref{eq:projection_operator} simplifies to $P_{ij}|_{\rm nr}=\delta_{ij}+\mathcal{O}(k^2/m^2)$. Upon making these approximations, we arrive at the following free Schr\"{o}dinger action for massive scalar, vector and tensor fields:
\begin{align}\label{eq:S_nr_free}
    \mathcal{S}^{\mathrm{eff}}_{\mathrm{free}} &= \int \mathrm{d}^4x\,\Biggl\{\frac{i}{2}\Tr\left[\bPsi^{\dagger}\dot{\bPsi}\right] + \mathrm{h.c.}\nonumber\\
    &\qquad\qquad\qquad\qquad -\frac{1}{2m} \Tr[\nabla\bPsi^{\dagger}\cdot\nabla\bPsi]\Biggr\}.
\end{align}
For the case of spin-$0$, $\bPsi$ carries no spatial index, while for the case of spin-$1$ and spin-$2$, it carries one and two spatial indices respectively. We shall refer to their components as 
\begin{align}
    &\psi=[\bPsi]\qquad&\textrm{spin-$0$}\,,\nonumber\\
    &\psi_i=[\bPsi]_i\qquad&\textrm{spin-$1$}\,,\\
    &\psi_{ij}=[\bPsi]_{ij}\qquad&\textrm{spin-$2$}\,.\nonumber
\end{align}
Note that for the spin-$1$ and spin-$2$ cases, $\Tr[\bPsi^\dagger\bPsi]=\psi_i^\dagger\psi_i$ and $\Tr[\bPsi^\dagger\bPsi]=\psi_{ij}^{\dagger}\psi_{ij}$ respectively, where  summation over indices in implicit.\footnote{Throughout the text, we implicitly assume summation over repeated indices, unless mentioned otherwise.} Also for the massive spin-$2$ case, we have $\Tr[\bPsi] = \mathcal{O}(k^2\bPsi/m^2)$ and therefore can be neglected in the non-relativistic limit. Furthermore, we obtain the following general structure for the non-relativistic energy momentum tensor
\begin{align}\label{eqn:emt_gen}
    \mathcal{T}_{00} &= m\,\Tr\left[\bPsi^{\dagger}\bPsi\right] + \mathcal{O}\left(\frac{k^2}{m}\bPsi^{\dagger}\bPsi\right),\nonumber\\
    \mathcal{T}_{0i} &= \mathcal{O}\left(k\,\bPsi^{\dagger}\bPsi\right),\nonumber\\
    \mathcal{T}_{ij} &= \mathcal{O}\left(\frac{k^2}{m}\bPsi^{\dagger}\bPsi\right).
\end{align}
With this, we can see from \eqref{eq:interaction_physical} that there are no gravitons (gravitational waves) produced in the non-relativistic limit since the source term for $h_{ij}$ is $\mathcal{O}(k^2\bPsi^{\dagger}\bPsi/m)$ for all the three cases. From \eqref{eq:h_constraints}, it is also clear that the vector constraint $h_{0i}$ is not sourced at the leading order in either of the three cases. The only constraint that survives is the Newtonian potential $\Phi$, which is determined by
\begin{align}\label{eq:Phi_NR}
    \nabla^2\Phi = \frac{m}{2\mpl^2}\,\Tr[\bPsi^\dagger \bPsi] + \mathcal{O}\left(\frac{k^2}{m}\bPsi^{\dagger}\bPsi\right).
\end{align}
Note that the trace ${\rm Tr}[h_{ij}]$ is equal to twice the Newtonian potential, ${\rm Tr}[h_{ij}] = 2\Phi$. Putting it all together, using \eqref{eq:S_nr_free}, \eqref{eqn:emt_gen}, and \eqref{eq:Phi_NR}, we get the following Schr\"{o}dinger Poisson action
\begin{align}\label{eq:master_action}
    \mathcal{S}^{\mathrm{eff}}_{nr} &= \int \mathrm{d}^4x\,\Bigl[\frac{i}{2}\Tr\left[\bPsi^{\dagger}\dot{\bPsi}\right] + \mathrm{c.c.} - \frac{1}{2m} \Tr[\nabla\bPsi^{\dagger}\cdot\nabla\bPsi]\nonumber\\
    &\qquad\qquad\quad + \mpl^2\,\Phi\nabla^2\Phi - m\,\Phi\,\Tr[\bPsi^{\dagger}\bPsi]\Bigr].
\end{align}
The corresponding equation of motion is the Schr\"{o}dinger-Poisson system
\begin{align}\label{eq:SP_general}
    i\frac{\partial}{\partial t}\bPsi &= -\frac{1}{2m}\nabla^2\bPsi + m\,\Phi\,\bPsi\,,\nonumber\\
    \nabla^2\Phi &= \frac{m}{2\mpl^2}\,\Tr[\bPsi^\dagger \bPsi].
\end{align}
Extension to an FLRW universe can be achieved via $\nabla\rightarrow \nabla/a$ and $\partial/\partial t\rightarrow \partial/\partial t+3H/2$ where $H=\dot{a}/a$.\footnote{For the spin-2 case, one needs to ensure that the choice of FLRW background is consistent within the bi-gravity + matter theory. While some aspects of this are discussed in the appendix, we leave a detailed exploration to future work.} 
\subsection{Conservation Laws}
We now highlight various symmetries of the non-relativistic effective theory~\eqref{eq:master_action}. These symmetries will be helpful in understanding the space of soliton solutions. 

First, the action in invariant under $\bPsi({\bf x}) \rightarrow \mathcal{M}(\bR)\,\bPsi(\tilde{\bf x})$, where $\bR$ is a rotation matrix and $\tilde{\bf x} = \bR^{-1}{\bf x}$. For scalars, $\mathcal{M}(\bR)\,\bPsi(\tilde{\bf x}) = \psi(\tilde{\bf x})$; for vectors $\mathcal{M}(\bR)\,\bPsi(\tilde{\bf x}) = R_{ij}\psi_{j}(\tilde{\bf x})$; and for tensors $\mathcal{M}(\bR)\,\bPsi(\tilde{\bf x}) = R_{ik}\psi_{kl}(\tilde{\bf x})R_{lj}$. 
The conserved charge density associated with this, namely the total angular momentum density is the following
\begin{align}\label{eq:total_ang_mom}
    J_k &= s\,\Re\left(i\,\varepsilon_{ijk}[\bPsi\bPsi^{\dagger}]_{ij}\right) +  \Re\left(i\,\varepsilon_{ijk}\mathrm{Tr}[\bPsi^{\dagger}\partial_i\bPsi]x^j\right)
\end{align}
where we identify
\begin{align}\label{eq:spin__ang_mom}
S_k&=s\,\Re\left(i\,\varepsilon_{ijk}[\bPsi\bPsi^{\dagger}]_{ij}\right),\\
L_k&=\Re\left(i\,\varepsilon_{ijk}\mathrm{Tr}[\bPsi^{\dagger}\partial_i\bPsi]x^j\right)\nonumber
\end{align}
as the intrinsic spin and orbital angular momentum density respectively.\footnote{Note that the total spin, and total orbital angular momentum are independently conserved.} 
In the above expressions, $s = \{0,1,2\}$ for spin-$0$ (scalar), spin-$1$  (vector), and spin-$2$ (tensor) cases respectively, and $\varepsilon_{a}$ are the totally anti-symmetric matrices (generators of rotations). 
Also, $[..]_{ij}$ means the matrix obtained by taking a tensor product of the elements within. For the vector case $[\bPsi\bPsi^{\dagger}]_{ij} = \psi_{i}\psi^{\dagger}_{j}$, while for the tensor case $[\bPsi\bPsi^{\dagger}]_{ij} = \psi_{ik}\psi^{\dagger}_{kj}$. For the scalar case this is zero.  \\

The effective non-relativistic action \eqref{eq:master_action} has a global $U(1)$ symmetry  ($\bPsi\rightarrow \bPsi e^{i\alpha}$) leading to a conserved particle number 
\begin{align}\label{eq:total_number}
    N=\int d^3 x {\rm Tr}[\bPsi^\dagger \bPsi]\,.
\end{align}
\noindent Along with rotational invariance, the usual time-translation invariance\footnote{There will also be linear momentum associated with invariance under spatial translations. This however will not be of any direct use for us.} of \eqref{eq:master_action} yields a conserved energy:
\begin{align}\label{eq:energy}
    E &= \int\mathrm{d}^3x\Bigl[\frac{1}{2m}\,\Tr[\nabla \bPsi^{\dagger}\cdot\nabla \bPsi] \nonumber\\
    &+ \frac{m^2}{4\mpl^2}\Tr[\bPsi^{\dagger}\bPsi]\int \frac{\mathrm{d}^3y}{4\pi|{\bf x}-{\bf y}|}\Tr[\bPsi^{\dagger}({\bf y})\bPsi(\bf y)]\Bigr]
\end{align}
\noindent There are additional conserved charges, which become apparent when we decompose our field $\bPsi$ into a polarization bases, which we turn to next.
\subsection{Decomposition into polarization basis}
A massive spin-$s$ field admits $2s+1$ spin multiplicity states in some particular direction $\hat{n}$, labelled by $\lambda \in \{-s,...,s\}$. These states are characterized by the set $\{\bepsilon^{(\lambda)}_{s,\hat{n}}\}$, such that upon substituting a plane wave $\bPsi^{(\lambda)}= V^{-1/2} e^{i{\bf k}\cdot{\bf x}}\bepsilon^{(\lambda)}_{s,\hat{n}}$ in \eqref{eq:spin__ang_mom}, we get 
\begin{align}
    \hat{n}\cdot\bS\left(\bPsi^{(\lambda)}\right) =\hat{n}\cdot\bS\left(\bepsilon^{(\lambda)}_{s,\hat{n}}\right) = \frac{\lambda}{V} \qquad \forall \lambda \in \{-s,..s\}
\end{align}
where $V$ is a spatial volume. For explicit forms of $\bepsilon^{(\lambda)}_{s,\hat{n}}$, see~\eqref{eq:basis_vectors} and~\eqref{eq:basis_tensors} ahead for spin-$1$ and spin-$2$ cases, where we work with $\hat{n}=\hat{z}$ without loss of generality. The set $\{\bepsilon^{(\lambda)}_{s,\hat{n}}\}$ is orthogonal and complete in the sense:
\begin{align}\label{eq:orthonormality_condition}
   \Tr[{\bepsilon}^{(\lambda)\;\dagger}_{s,\hat{n}}{\bepsilon}^{(\lambda')}_{s,\hat{n}}] = \delta_{\lambda,\lambda'},\quad\sum_{\lambda}\left[{\bepsilon}^{(\lambda)}_{s,\hat{n}}\;{\bepsilon}^{(\lambda)\dagger}_{s,\hat{n}}\right]_{ij}\propto\delta_{ij}
   .
\end{align}
Using these $\{\bepsilon_{s,\hat{n}}^{(\lambda)}\}$, the field $\bPsi$ admits the following decomposition
\begin{align}\label{bPsi_pol_eig}
    \bPsi({\bf x},t) = \sum_{\lambda}\psi^{(\lambda)}_s({\bf x},t){\bepsilon}^{(\lambda)}_{s,\hat{n}}\,,
\end{align}
where $\psi^{(\lambda)}_s$ is the field with polarization $\lambda$ in the $\hat{n}$ direction. In terms of these different polarized fields $\psi^{(\lambda)}_s$, the action has the following form
\begin{align}\label{eq:master_action_2}
    \mathcal{S}^{\mathrm{eff}}_{nr} &=\sum_\lambda\int \mathrm{d}^4x\,\Bigl[\frac{i}{2}\psi^{(\lambda)\;\dagger}_s\dot{\psi}^{(\lambda)}_s + \mathrm{c.c.}\nonumber\\
    &\qquad\qquad -\frac{1}{2m} \nabla\psi^{(\lambda)\;\dagger}_s\cdot\nabla\psi^{(\lambda)}_s + m_{\rm pl}^2\,\Phi\nabla^2\Phi\nonumber\\
    &\qquad\qquad\qquad\quad - m\,\Phi\,\psi^{(\lambda)\;\dagger}_s\psi^{(\lambda)}_s\Bigr].
\end{align}
Correspondingly, the equation of motion is the Schr\"{o}dinger Poisson system where we have a set of Schr\"{o}dinger field equations for each $\psi^{(\lambda)}_s$, plus the Newtonian Gauss' law 
\begin{align}\label{eq:SP_general_2}
    i\frac{\partial}{\partial t}\psi^{(\lambda)}_s &= -\frac{1}{2m}\nabla^2\psi^{(\lambda)}_s + m\,\Phi\,\psi^{(\lambda)}_s\nonumber\\
    \nabla^2\Phi &= \frac{m}{2\mpl^2}\sum_{\lambda}\psi^{(\lambda)\;\dagger}_s\psi^{(\lambda)}_s.
\end{align}

The orbital and spin angular momentum densities~\eqref{eq:total_ang_mom}, in terms of the $\psi_s^{(\lambda)}$ are
\begin{align}\label{eq:total_ang_mom_2}
    S_k &= s\,\sum_{\lambda,\lambda'}\Re\left(i\,\psi_s^{(\lambda')\,\dagger}\psi^{(\lambda)}_{s}\,\varepsilon_{ijk}[{\bepsilon}^{(\lambda)}_{s,\hat{n}}{\bepsilon}^{(\lambda')\;\dagger}_{s,\hat{n}}]_{ij}\right)\nonumber\\
    L_k &= \sum_{\lambda}\Re\left(i\,\psi_s^{(\lambda)\,\dagger}\varepsilon_{ijk}\,\partial_i\psi_s^{(\lambda)}\,x^j\right).
\end{align}

Action ~\eqref{eq:master_action_2}, written in terms of the $\psi^{(\lambda)}_s$, is helpful to identify another set of symmetries. We have global U(1) invariance for {\it each} of the $2s+1$ degrees of freedom, giving $2s+1$ conserved particle numbers
\begin{align}\label{eq:number_each}
    N^{(\lambda)} = \int \mathrm{d}^3x\,\psi^{(\lambda)\;\dagger}_s\psi^{(\lambda)}_s,
\end{align}
where $N = \sum_{\lambda}N^{(\lambda)}$. The fact that each of the polarized fields has an associated conserved particle number, will become important to physically understand superpositions of extremally polarized solitons in order to form fractionally polarized ones.\\

Finally, the conserved energy~\eqref{eq:energy} written in terms of $\psi^{(\lambda)}_{s}$ is
\begin{align}\label{eq:energy2}
    E 
    &= \sum_\lambda \Bigg\{\int\mathrm{d}^3x\Bigl[\frac{1}{2m} \nabla\psi^{(\lambda)\;\dagger}_s\cdot\nabla\psi^{(\lambda)}_s\nonumber\\
    &+ \frac{m^2}{4\mpl^2}\psi^{(\lambda)\;\dagger}_s\psi^{(\lambda)}_s\sum_{\lambda'}\int \frac{\mathrm{d}^3y}{4\pi|{\bf x}-{\bf y}|}\,\psi^{(\lambda')\;\dagger}_s({\bf y})\psi^{(\lambda')}_s({\bf y})\Bigr]\Bigg\}.
\end{align}

On a conceptual note,  the non-relativistic action \eqref{eq:master_action_2} cannot distinguish between (i) a set of $2s+1$ (spatial-)scalars (ii) a spin-$s$ field with $2s+1$ spin multiplicity degrees of freedom. While the set of (spatial-)scalars will have no spin angular momentum, they will still have a conserved ``isospin" \eqref{eq:total_ang_mom_2}.\footnote{The phenomenology with multiple scalar fields, but with different masses, was explored in~\cite{Eby:2020eas}.} The non-relativistic theory cannot distinguish between the two cases using gravitational physics alone. This equivalence can be broken when we include relativistic corrections.

\section{Polarized Solitons}\label{sec:grav_bound}
Making use of the polarization basis~\eqref{bPsi_pol_eig}, we classify different (lowest energy) solitons based on their spin multiplicities. We first discuss `extremally polarized' solitons (composed of identically polarized plane waves), where only one of the $2s+1$ polarized fields $\psi_s^{(\lambda)}$ is non-zero. Thereafter, we discuss fractionally polarized solitons obtained by linear superpositions of extremally polarized ones.
\subsection{Extremally polarized solitons}
\begin{figure}[t]
\begin{center}
\includegraphics[width=0.45\textwidth]{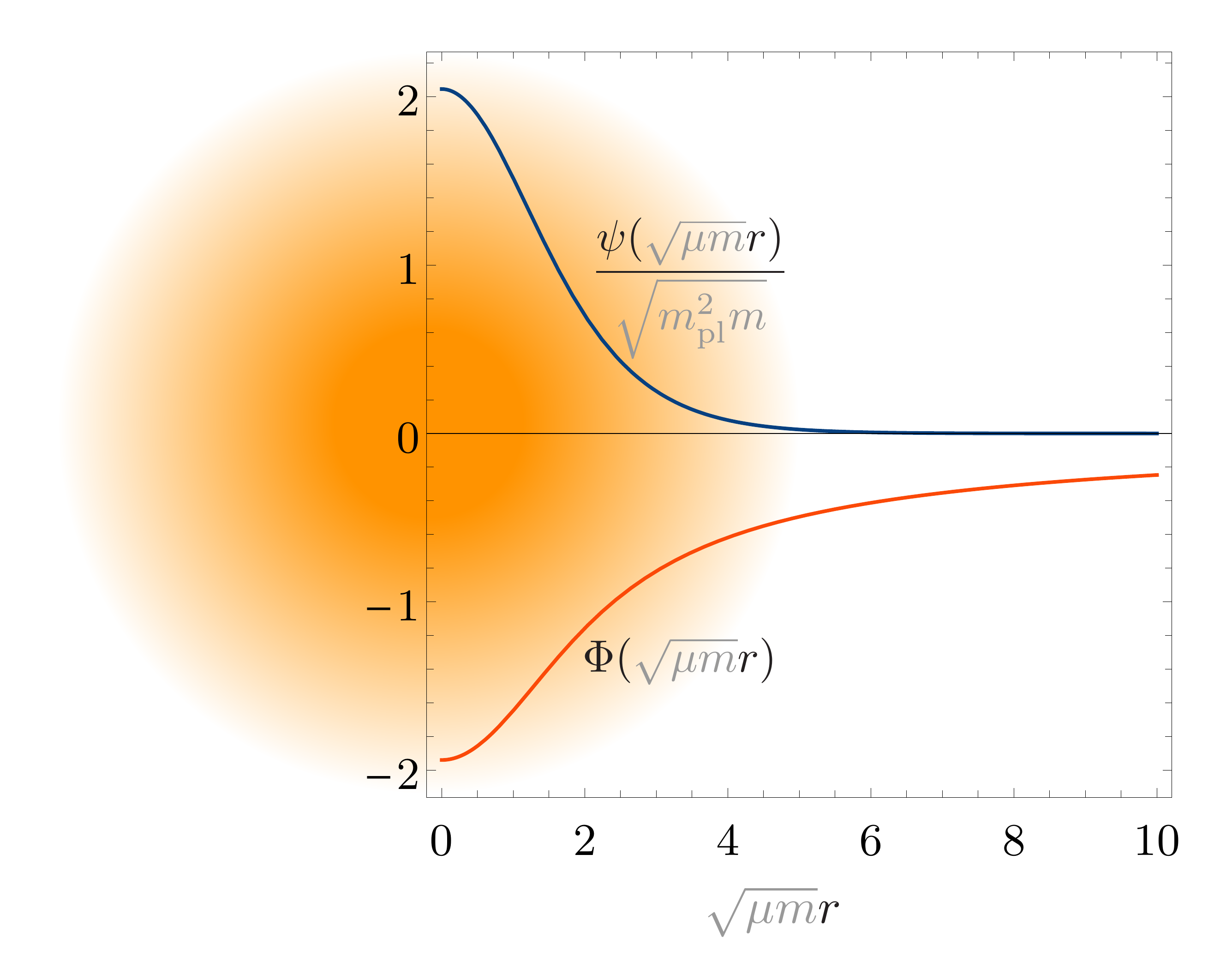}
\end{center}
\caption{The ``universal" non-relativistic field profile and gravitational potential for the soliton solutions. Note that both profiles must be multiplied by $\mu/m$ to get the correct solution for each $\mu/m$.} 
\label{fig:profile}
\end{figure}

With only one of the polarization fields non-zero, we can assume the following ansatz \begin{align}\label{eq:psi_lambda_ansatz}
    \psi^{(\lambda)}_s({\bf x},t) = \psi({\bf x})\,e^{i\mu t}\,.
\end{align}
That is, $\psi^{(\lambda)}\ne0$ for a particular $\lambda$, and zero otherwise. Here $\mu = const.$ can be thought of as the `chemical potential'. In this case, we have the usual (scalar) Schr\"{o}dinger-Poisson system
\begin{align}\label{psi_equation}
    -\mu\,\psi &= -\frac{1}{2m}\nabla^2\psi + m\,\Phi\,\psi\nonumber\\
    \nabla^2\Phi &= \frac{m}{2\mpl^2}\,|\psi({\bf x})|^{2},
\end{align}
along with the following energy~\eqref{eq:energy}:
\begin{align}
    E &= \int\mathrm{d}^3x\Bigl[\frac{1}{2m} \nabla\psi^{\dagger}\cdot\nabla\psi\nonumber\\
    &\qquad\qquad\qquad + \frac{m^2}{4\mpl^2}\psi^{\dagger}\psi\int \frac{\mathrm{d}^3y}{4\pi|{\bf x}-{\bf y}|}\,\psi^{\dagger}({\bf y})\psi({\bf y})\Bigr].
\end{align}
The lowest energy solution to eq. \eqref{psi_equation} constitutes the well known, spherically symmetric, scalar soliton configuration~\cite{PhysRev.187.1767}. The unique profile (up to scaling of the amplitudes by $\mu/m$) is shown in Fig.~\ref{fig:profile}.\\
\begin{figure*}[t]
\begin{center}
\includegraphics[width=0.9\textwidth]{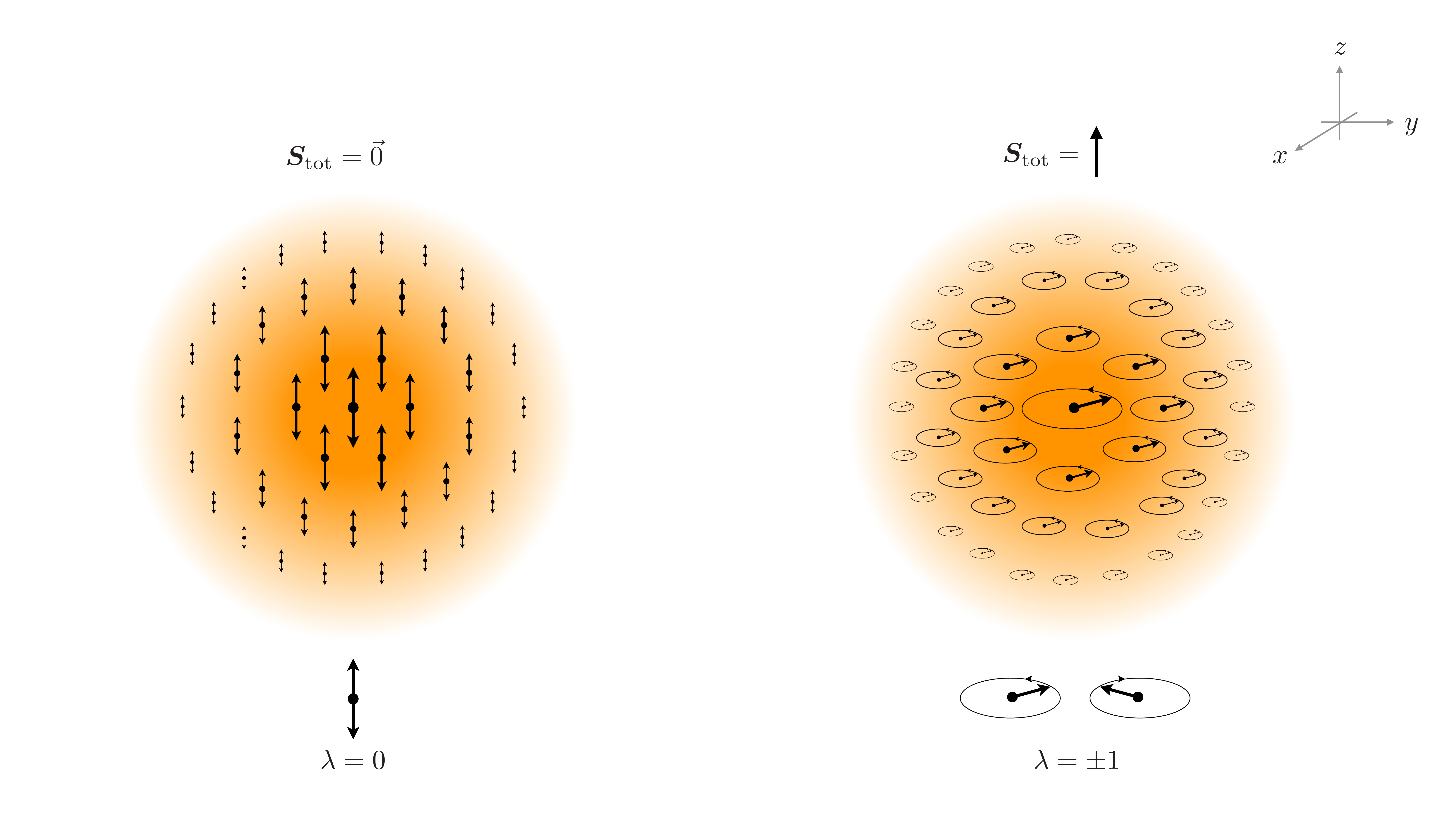}
\end{center}
\caption{A visualization of the two {\it distinct} extremally polarized vector solitons. The left soliton has vanishing spin density ($\lambda=0$), and $\bW$ is oscillating along the $z$-axis. The right soliton has a spin density $\bS = \lambda |\psi|^2\hat{z}$ with $\lambda=1$. The big arrows inside the soliton represent the direction of the field $\bW$, while the little arrows on the circles represent their motion in time. The total spin $|\bS_{\rm tot}|=\lambda M/m\approx60.7\,\lambda\,  (\mpl/m)^2(\mu/m)^{1/2}$, where $M$ is the total mass of the soliton.} 
\label{fig:v_solitons}
\end{figure*}
\begin{figure*}[t]
\begin{center}
\includegraphics[width=0.9\textwidth]{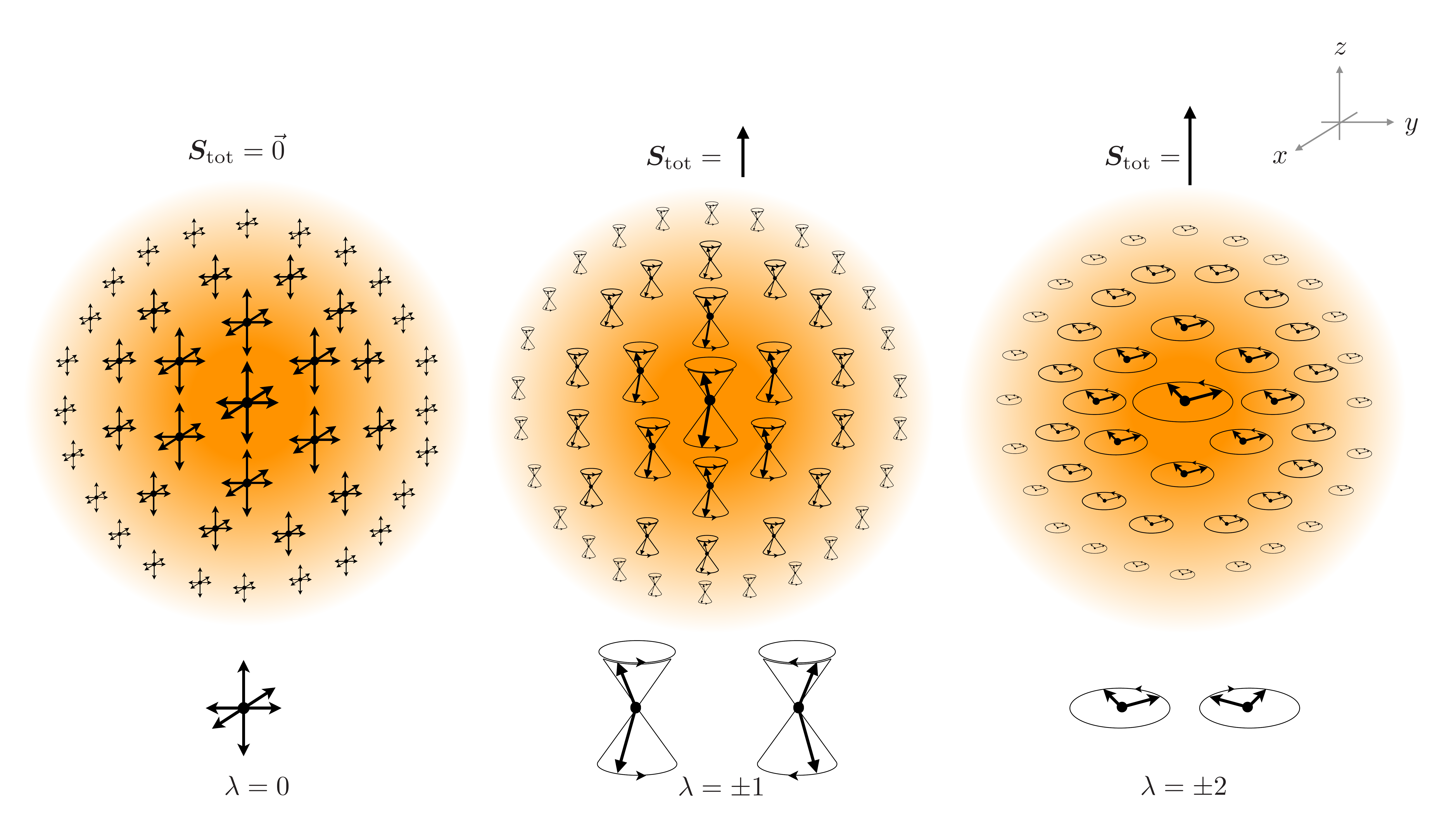}
\caption{The three {\it distinct} extremally polarized tensor solitons. For visualization, we plot the eigenvectors of the traceless, $3\times 3$ matrix $\bH^{(\lambda)}$ representing the polarized solitons in the massive spin-$2$ field. The eignevectors are scaled by their corresponding eigenvalues. The leftmost soliton has a vanishing spin density, with each eigenvector (along the co-ordinate axes) of $\bH^{(\lambda)}$ oscillating in phase. The middle soliton has a spin density $\bS=\lambda|\psi|^2\hat{z}$ with $\lambda=1$. The rightmost soliton has a spin density $\bS = \lambda|\psi|^2\hat{z}$ with $\lambda=2$. The total spin $|\bS_{\rm tot}|=\lambda M/m\approx60.7\,\lambda\,  (\mpl/m)^2(\mu/m)^{1/2}$ which can be macroscopically large for $\lambda\ne0$.} 
\label{fig:t_solitons}
\end{center}
\end{figure*}

Since $\psi$ is real and $\mu$ spatially independent, the orbital angular momentum in~\eqref{eq:total_ang_mom_2} vanishes:
\begin{align}\label{eq:L_solitons}
    \bm{L} = \Re\left(i\,\psi\nabla\psi\times\bm{r}\right) = 0\,.
\end{align}
For higher-spin fields, the spin angular momentum density can be non-vanishing. For the ansatz~\eqref{eq:psi_lambda_ansatz}, it is equal to 
\begin{align}\label{eq:spindensity_solitons_spin1}
    \bS = \lambda\psi^2\hat{n}
    \,,\qquad \lambda = \{-s,...,0,...,s\}.
\end{align}
This reflects the extremally polarized nature of the solitons. For each $\lambda$, this is a coherent collection of plane waves, all polarized along the $\hat{n}$ direction. Explicitly, the total spin angular momentum 
\begin{align}
{\bS}_{\rm tot}=\int d^3x \bS=\lambda \int d^3x |\psi|^2 \hat{n}=\lambda N\hat{n}.
\end{align}
where $N$ is the total particle number in the soliton.\\

For extremally polarized solitons, the particle number, energy and spin-angular momentum are 
\begin{align}
\label{eq:Num_NES}
    N&=\frac{M}{m}\approx60.7\frac{\mpl^2}{m^2}\left(\frac{\mu}{m}\right)^{\!1/2}\,,\\
    E&\approx-19.2\frac{\mpl^2}{m}\left(\frac{\mu}{m}\right)^{\!3/2}\,,\\ 
    \bS_{\rm tot}&\approx\lambda \times60.7\frac{\mpl^2}{m^2}\left(\frac{\mu}{m}\right)^{\!1/2}\hat{n}\,,
\end{align}
where $M$ is the total mass of the soliton. The numerical co-efficients are obtained from the universal profile shown in Fig.~\ref{fig:profile}. Heuristically $(\mu/m)^{1/2}\sim (1/mL)$ where $L$ is the characteristic size of the soliton, and $M \sim (\mpl/m)^2/L$. Since in the non-relativistic regime $\mu/m\ll 1$, we expect the maximal values of the above quantities~\eqref{eq:Num_NES} to be bounded from above by the case $\mu\sim m$. Significant deviations from the above expression can be expected as one approaches this limit \cite{Salehian:2021khb}.\\

It is also worth noting that these solitons are perfectly virialized, $E_{\rm kin}/E_{\rm pot}=-1/2$, where $E_{\rm kin}$ is the term containing gradients, and $E_{\rm pot}$ is the other term (gravitational potential energy) in \eqref{eq:energy}.
\subsubsection{Spin-$0$}
For the case of spin-$0$, we have the following real field solution for $\phi (= \bm{\mathcal{F}})$ in~\eqref{eq:F_decompose}
\begin{align}
    \phi({\bf x},t) = \frac{\sqrt{2}\psi({\bf x})}{
    \sqrt{m}}\cos\omega t\,,
\end{align}
where $\omega \equiv m-\mu$, and there is of-course no intrinsic spin angular momentum.
\subsubsection{Spin-$1$}
For the massive spin-$1$ case, we have three distinct states corresponding to $\pm 1$ and $0$ polarizations, which, for $\hat{n}=\hat{z}$, are conveniently represented by the following orthonormal set of vectors:
\begin{align}\label{eq:basis_vectors}
    \bepsilon^{(\pm 1)}_{1,\hat{z}} = \frac{1}{\sqrt{2}}
    \begin{pmatrix}
        1\\
        \pm i\\
        0
    \end{pmatrix}\;;&\qquad
    \bepsilon^{(0)}_{1,\hat{z}} =
    \begin{pmatrix}
        0\\
        0\\
        1
    \end{pmatrix}.
\end{align}
satisfying~\eqref{eq:orthonormality_condition}. For $\bPsi^{(\lambda)}=\psi e^{i\mu t}\bepsilon_{1,\hat{z}}^{(\lambda)}$, we have $\bS=\lambda|\psi|^2\hat{z}$ where $\lambda=0,\pm 1$. Extremally polarized solitons in terms of the real-valued vector field ${\bW} (= \bm{\mathcal{F}})$ in~\eqref{eq:F_decompose} are
\begin{align}\label{eq:W_solutions}
    {\bW}^{(\pm 1)}({\bf x},t) &= \frac{\psi({\bf x})}{\sqrt{m}}
    \begin{pmatrix}
        \cos\omega t\\
        \pm\sin\omega t\\
        0
    \end{pmatrix}\,,\nonumber\\
    \bW^{(0)}({\bf x},t) &= \frac{\sqrt{2}\,\psi({\bf x})}{\sqrt{m}}
    \begin{pmatrix}
        0\\
        0\\
        \cos\omega t 
    \end{pmatrix}.
\end{align}
In fig.~\ref{fig:v_solitons} we show these extremally polarized solitons.\\

We note that in~\cite{Adshead:2021kvl}, the authors provide ground state solitons $\bPsi = \psi(r)e^{i\mu t}\{w_x,w_y,w_z\}$ where $w_i$ are components of a complex unit vector. For the extremally polarized cases, this corresponds to the choices~\eqref{eq:basis_vectors} for their $w_i$. Our focus on the spin aspect of fields dictated this choice, which naturally leads to extremally polarized solitons. More general solitons with arbitrary $w_i$ are discussed in~\ref{sec:frac_pol}.\\
\subsubsection{Spin-2}
For the massive spin-$2$ case we have 5 polarization states  corresponding to spin multiplicities $\pm 2$, $\pm 1$, and $0$. Again, for $\hat{n}=\hat{z}$, these are represented by the following orthonormal (and trace free) set of tensors\footnote{These can be obtained through tensor products of the spin-1 polarization vectors: $\epsilon^{(\pm 2)}_{2,\hat{z}} = \epsilon^{(\pm 1)}_{1,\hat{z}}\odot\epsilon^{(\pm 1)}_{1,\hat{z}}$; $\epsilon^{(\pm 1)}_{2,\hat{z}} = (\epsilon^{(\pm 1)}_{1,\hat{z}}\odot\epsilon^{(0)}_{1,\hat{z}} + \epsilon^{(0)}_{1,\hat{z}}\odot\epsilon^{(\pm 1)}_{1,\hat{z}})/\sqrt{2}$; and $\epsilon^{(0)}_{2,\hat{z}} = (2\epsilon^{(0)}_{1,\hat{z}}\odot\epsilon^{(0)}_{1,\hat{z}} - \epsilon^{(+1)}_{1,\hat{z}}\odot\epsilon^{(-1)}_{1,\hat{z}} - \epsilon^{(- 1)}_{1,\hat{z}}\odot\epsilon^{(+1)}_{1,\hat{z}})/\sqrt{6}$.}
\begin{align}\label{eq:basis_tensors}
    \bepsilon^{(\pm 2)}_{2,\hat{z}} &= \dfrac{1}{2}\begin{pmatrix}
    1 && \pm i && 0\\
    \pm i && -1 && 0\\
    0 && 0 && 0
    \end{pmatrix}\nonumber\\
    \bepsilon^{(\pm 1)}_{2,\hat{z}} &= \dfrac{1}{2}\begin{pmatrix}
    0 && 0 && 1\\
    0 && 0 && \pm i\\
    1 && \pm i && 0
    \end{pmatrix}\nonumber\\
    \bepsilon^{(0)}_{2,\hat{z}} &= \dfrac{1}{\sqrt{6}}\begin{pmatrix}
    -1 && 0 && 0\\
    0 && -1 && 0\\
    0 && 0 && 2
    \end{pmatrix}
\end{align}
 For $\bPsi^{(\lambda)}=\psi e^{i\mu t} \bepsilon_{2,\hat{z}}^{(\lambda)}$, we have the spin density $\bS=\lambda|\psi|^2\hat{z}$ where $\lambda=0,\pm 1,\pm 2$.

The five extremally polarized solitons in the real-valued trace-free tensor field are:
\begin{align}\label{eq:H_solutions}
    {\bH}^{(\pm 2)}({\bf x},t) &= \frac{\psi({\bf x})}{\sqrt{2m}}
    \begin{pmatrix}
        \cos\,\omega t & \pm\sin\,\omega t & 0\\
        \pm\sin\,\omega t & -\cos\,\omega t & 0\\
        0 & 0 & 0
    \end{pmatrix}\nonumber\\
    {\bH}^{(\pm 1)}({\bf x},t) &= \frac{\psi({\bf x})}{\sqrt{2m}}
    \begin{pmatrix}
        0 & 0 & \cos\,\omega t\\
        0 & 0 & \pm\sin\,\omega t\\
        \cos\,\omega t & \pm\sin\,\omega t & 0
    \end{pmatrix}\nonumber\\
    {\bH}^{(0)}({\bf x},t) &= \frac{\psi({\bf x})}{\sqrt{3m}}
    \begin{pmatrix}
        -1 && 0 && 0\\
    0 && -1 && 0\\
    0 && 0 && 2
    \end{pmatrix}\,\cos\,\omega t,
\end{align}
Fig.~\ref{fig:t_solitons} shows these extremally polarized solitons (with spin density along the $z$ axis). 
\subsection{Fractionally polarized solitons}\label{sec:frac_pol}
Here we construct non-extremal polarized solitons, obtained through linear superpositions of the extremal ones. Since we have U(1) invariance for each polarization field $\psi_s^{(\lambda)}$, we can superpose them to form new solutions. That is to say that we can have
\begin{align}\label{eq:superposition_vec}
    \bPsi({\bf x},t) &= \psi({\bf x})\sum_{\lambda}c_{\lambda}e^{i(\mu t - \varphi_\lambda)}{\bepsilon}^{(\lambda)}_{s,\hat{n}}
\end{align}
with
\begin{align}\label{eq:c_normalize}
    \qquad\sum_{\lambda}c^2_{\lambda} = 1.
\end{align}
The reason to enforce~\eqref{eq:c_normalize} is to ensure that we have the same particle number density as in the case of extremally polarized solitons (and hence also the same energy), which in turn guarantees that $\psi({\bf x})$ obeys the same Schr\"{o}dinger Poisson system~\eqref{psi_equation}.\\

The corresponding real valued counterpart is
\begin{align}\label{eq:F_general}
\bm{\mathcal{F}}({\bf x},t)=\sum_\lambda c_\lambda  \bm{\mathcal{F}}^{(\lambda)}({\bf x},t+\varphi_\lambda/\omega)\,,
\end{align}
where $\bm{\mathcal{F}}^{(\lambda)}$ are extremally polarized solitons (equal to ${\bW}^{(\lambda)}$ for spin-$1$ \eqref{eq:W_solutions}; ${\bH}^{(\lambda)}$ for spin-$2$ \eqref{eq:H_solutions}).\\

Similar to the case of extremally polarized solitons, the orbital angular momentum density, obtained by substituting \eqref{eq:superposition_vec} into \eqref{eq:total_ang_mom_2} is zero. On the other hand the spin density is 
\begin{align}
S_k=s|\psi({\bf x})|^2\sum_{\lambda\lambda'}\Re\left[ic_\lambda c_{\lambda'}e^{i(\varphi_\lambda-\varphi_{\lambda'})}\epsilon_{ijk}[\bepsilon_{s,\hat{n}}^{(\lambda')}\bepsilon_{s,\hat{n}}^{(\lambda)\dagger}]_{ij}\right].
\end{align}
The total spin need not be equal to $\lambda N$ where $\lambda \in \{-s,..,s\}$ for general $\{c_\lambda,\varphi_\lambda\}$. However, for a class of $\{c_\lambda,\varphi_\lambda\}$ which simply amount to rotations of extremally polarized solitons, we will again get $\lambda N$ for the total spin.\\

Superposing basis solutions (extremally polarized) to form new solutions, while keeping the total particle number fixed, is just taking different fractions of these extremally polarized solitons and putting them on top of each other (c.f. \eqref{eq:F_general}). The reason this is allowed is because there is a U(1) invariance within each polarization sector. From a phenomenological point of view, there could be extra spin induced interactions that favor same polarization states for a collection of particles in some region (e.g. Ising model). For such situations, it may be that there is a higher chance of extremally polarized solitons to form over fractionally polarized ones due to Bose-Einstein statistics. We leave such questions for future work.
\subsection{$(s/2,s/2)$ representation of polarized solitons}
There exists a simple understanding of the space of all polarized solitons in terms of the $(m,n)$ representation of the rotation group SO(3). Guided by the U(1) and SO(3) invariance of the general action \eqref{eq:master_action_2}, and the fact that there are $2s+1$ d.o.f. with $\lambda \rightarrow -\lambda$ obtainable via simple rotation, we can represent our full space of soliton solutions via $(s/2,s/2)$ representation of the SO(3) group. The representation is reducible, meaning it admits a total of $s+1$ SO(3) invariant sub-spaces, each containing the different (absolute value of-) polarizations $|\lambda| \in \{0,1,..s\}$. These are our {\it distinct} extremally polarized solitons.
These $s+1$ extremally polarized solitons form a basis, and can give rise to fractionally polarized solitons via appropriate superpositions (due to separate U(1) within each polarization sector). The full space of solutions is the product space of these $s+1$ subspaces, hosting a general soliton. A physical soliton spontaneously breaks the SO(3) invariance of the action~\eqref{eq:master_action}, (or equivalently~\eqref{eq:master_action_2}). Table~\ref{tab:table1} shows the representation of gravitationally bound polarized solitons in bosonic fields considered in this paper.

\begin{table}[b]
\caption{\label{tab:table1}%
$(m,n)$ representation of the gravitationally bound soliton states in non-relativistic integer spin field theories
}
\begin{tabular}{lcdr}
\textrm{Soliton states}&\qquad
\textrm{Massive field (spin)}&\\
\colrule
(0,0) & \textrm{scalar} (0) \\
(1/2,1/2) & \textrm{vector} (1) \\
(1,1) & \textrm{tensor} (2) \\
\end{tabular}
\end{table}

\begin{figure*}[t]
\begin{center}
\includegraphics[width=0.7\textwidth]{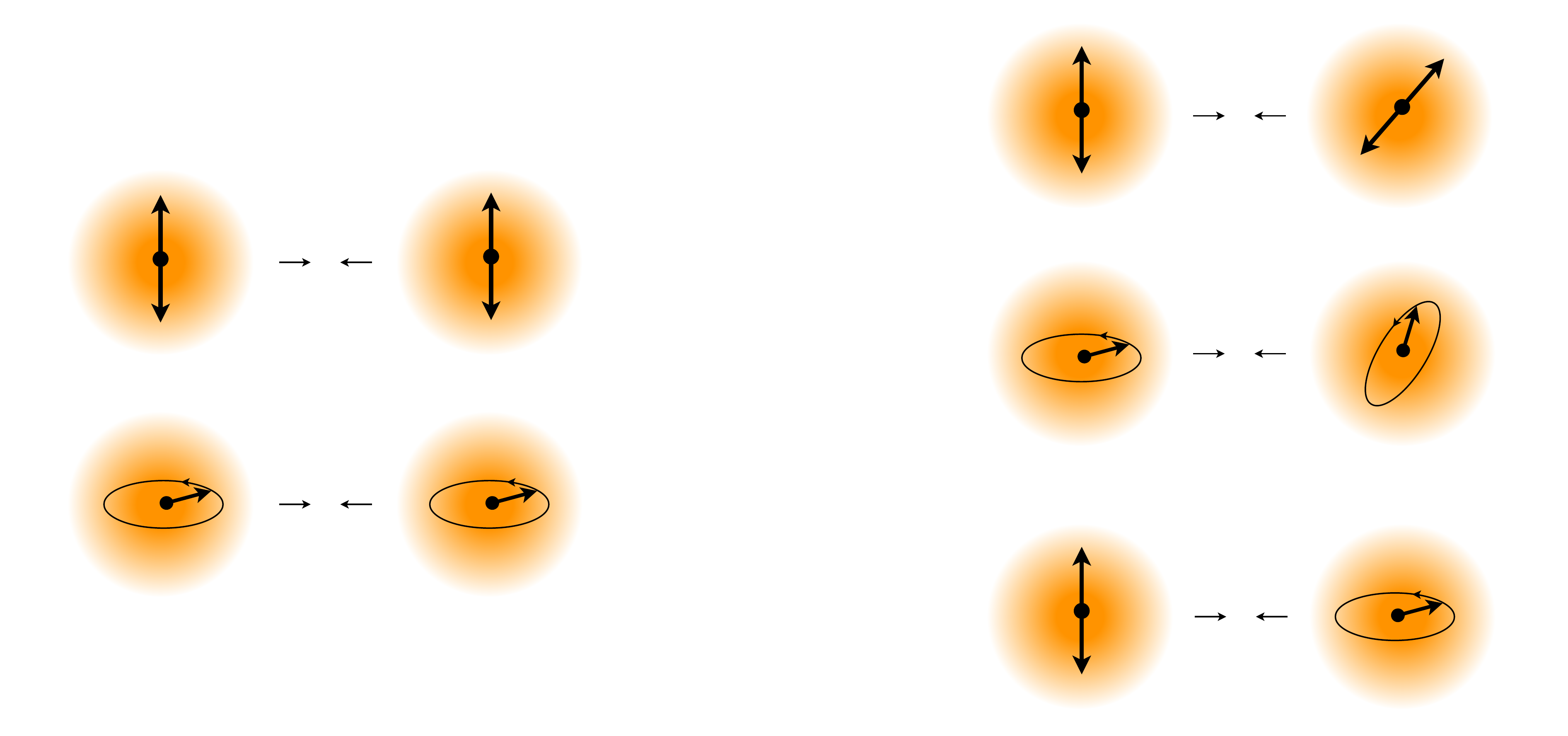}
\caption{The left panel shows collisions between vector solitons that can be replicated by solitons in a single scalar field. The right panel shows examples of collisions which cannot be replicated by solitons in a single scalar field.} 
\label{fig:Collision}
\end{center}
\end{figure*}
\subsection{Beyond ground-state solitons}
Spherically symmetric, single node solitons for the spin-$1$ and spin-$2$ fields are the `hedgehog'-like configurations, with Cartesian components \cite{Adshead:2021kvl,Aoki:2017ixz}\footnote{For relativistic hedgehogs in complex-valued Proca-fields, see \cite{Brito:2015pxa}.}
\begin{align}
    W_j({\bf x},t)&=f(r)\frac{x^j}{r}\cos \omega t\,,\nonumber\\
    H_{ij}({\bf x},t)&=g(r)\left(3\frac{x^i x^j}{r^2}-\delta_{ij}\right)\cos\omega t\,,
\end{align} 
where $f(0)=g(0)=0$. That is, there is a node in the profile at the origin. 
Both hedgehogs have higher energies (at a fixed particle number) compared to the ground state solitons discussed earlier, and have zero spin and orbital angular momentum. Explicitly, after fixing the particle number to be the same as the polarized solitons ($N^{s}_{\rm hh} = N$), we have $E^{s=1}_{\rm hh}\approx 0.33E$ and $E^{s=2}_{\rm hh}\approx 0.17 E$ where $E<0$ and $N$ are given in \eqref{eq:Num_NES}. Note that $\Delta E=E^{s}_{\rm hh}-E>0$. A linear stability analysis was provided in \cite{Aoki:2017ixz} to argue that the hedgehogs in spin-$2$ case are unstable and might transition to $p$-solitons. As with scalar solitons, excited configurations with additional nodes and orbital angular momentum might be possible with higher-spin fields, albeit with shorter lifetimes.\\ 

So far we have only allowed for spherically symmetric energy densities. It is possible to construct non-spherically symmetric configurations, such as domain walls, strings/vortices etc. \cite{Adshead:2021kvl};  the possible space of extended field configurations with higher-spin fields is likely to be quite rich. The full classification is beyond the scope of the present paper, but it is worth pursuing since it might provide new avenues to probe these higher-spin fields.

\section{Distinguishability \& probes of polarized solitons}\label{sec:distinguish}

Having shown that we have quite a rich space of soliton solutions, we briefly discuss some of the phenomenological implications. Alongside these implications, we address some conceptual questions: Can higher-spin solitons be distinguished from scalar solitons? Can solitons with different polarizations be distinguished using only gravitational interactions? 
\subsection{Gravitational interactions}

Let us consider collisions between solitons $A$ and $B$ in a spin-$s$ field. We show below that only if the two solitons differ by just an overall phase, can the collision be mimicked by two scalar solitons. Otherwise, in general, the higher-spin nature of the fields will leave an imprint in the observables related to the collision of the two solitons. 

For simplicity, let us consider a collision between two extremally polarized solitons, initially far away from each other, such that the field admits the following ansatz
\begin{align}
    \bPsi_s\Bigr|_{t=0} = 
    \psi_A\left({\bf x}+{\bf x}_0\right)\bepsilon_{s,\hat{n}}^{(\lambda)} +e^{i\theta} \psi_B\left({\bf x}-{\bf x}_{0}\right)\bepsilon_{s,\hat{n}'}^{(\lambda')}\,.
\end{align}
The corresponding initial number and current densities are
\begin{align}
    \mathcal{N}_s\Big|_{t=0}&=\psi_A^2+\psi_B^2+\psi_A\psi_B\left(e^{i\theta}\Tr[\bepsilon^{(\lambda')}_{s,\hat{n}'}\bepsilon^{(\lambda)\dagger}_{s,\hat{n}}] + c.c.\right)\nonumber\\
    {\bm{j}}_s\Big|_{t=0} &= \frac{i}{2m}\Bigl\{\left(e^{-i\theta} \Tr[\bepsilon^{(\lambda)}_{s,\hat{n}}\bepsilon^{(\lambda')\dagger}_{s,\hat{n}'}] - c.c.\right)\psi_B\nabla\psi_A\nonumber\\
    &\qquad +\left(e^{i\theta} \Tr[\bepsilon^{(\lambda')}_{s,\hat{n}'}\bepsilon^{(\lambda)\dagger}_{s,\hat{n}}] - c.c.\right)\psi_A\nabla\psi_B\Bigr\}.
\end{align}
For the scalar case ($s=0$, $\bepsilon_{0,\hat{n}}^{(\lambda)}=1$), we have
\begin{align}
    \mathcal{N}_0\Big|_{t=0}&=\psi_A^2+\psi_B^2+2\psi_A\psi_B\cos\theta\\
   \bm{j}_0\Big|_{t=0}&=\frac{\sin\theta}{m}
    \left(\psi_B\nabla\psi_A-\psi_A\nabla\psi_B\right).
\end{align}
Now to compare the collision scenario with the scalar case, we wish to start with equal number densities in the scalar {\it and} higher-spin cases. That is, $\mathcal{N}_s = \mathcal{N}_0$ at $t=0$. To achieve this condition we have two possibilities (1) $\Tr[\bepsilon^{(\lambda)}_{s,\hat{n}}\bepsilon^{(\lambda')\dagger}_{s,\hat{n}'}]\ne 1$ and real, with $\theta = \pi/2$; (2) $\Tr[\bepsilon^{(\lambda)}_{s,\hat{n}}\bepsilon^{(\lambda')\dagger}_{s,\hat{n}'}]= 1$, that is $\lambda=\lambda'$ and $\hat{n}=\hat{n}'$. For possibility (1), ${\bm{j}}_s = \vec{0}$ while ${\bm{j}}_0 \neq \vec{0}$ at $t=0$, so  the collision will proceed differently in the scalar vs. the higher-spin case. Specifically, the densities, and hence the gravitational potential, will also evolve differently in the two cases. Possibility (2) essentially reduces to the scalar case as expected. This result is summarized in Fig.~\ref{fig:Collision} (as an example for spin-$1$ fields). \\

Thus, using collisions we can tell whether two solitons in a given spin-$s$ field had the same spin density or not. This would not be possible with individual solitons in isolation using the gravitational potential alone. We leave a detailed analysis of how exactly the potential evolves and the associated phenomenological avenues such as motion of test particles in this dynamical potential \cite{Boskovic:2018rub} or merger rates of solitons \cite{Mocz:2017wlg} for future work.\footnote{With R. Karur and P. Mocz, we are investigating soliton interactions using numerical simulations of the multi-component SP system.} 

More generally, many existing studies carried out for scalar dark matter, can be repeated for the case of higher-spin dark matter including the spectra of associated cosmological perturbations, soliton formation mechanisms \cite{Amin:2019ums}, halo formation \cite{Mocz:2017wlg}, dynamical friction \cite{Lancaster:2019mde}, generation of gravitational waves \cite{Helfer:2018vtq}, time delays caused in pulsar timing arrays by time-dependent pressures and energy densities \cite{Khmelnitsky:2013lxt}, black-hole superradiance and `gravitational atoms'\cite{Brito:2015oca}, transient vortices~\cite{Hui:2020hbq} etc. Some of these have already been extended to higher-spin fields, see for example, \cite{Cembranos:2016ugq,PhysRevD.96.035019,Baumann:2019eav,Armaleo:2020yml}. 
\subsection{Non-gravitational interactions}
One can envision many scenarios where the higher-spin fields couple to the Standard Model such that it opens up channels for polarized solitons to be detectable. One possibility is to have some high energy scale(s) $g_{\mathcal{F}\gamma}^{-1}$, such that in the low-energy effective theory we have:
\begin{align}
    \mathcal{L}_{int}
    & \sim 
    \begin{cases} 
    g^2_{W\gamma}W_\mu W^\mu\,F_{\alpha\beta}\tilde{F}^{\alpha\beta} & {\rm spin-}1 \\ 
    g^2_{H\gamma}(H_{\mu\nu}H^{\mu\nu}-H^2)\,F_{\alpha\beta}\tilde{F}^{\alpha\beta} & {\rm spin-}2
    \end{cases}
    \;.
\end{align}
In the non-relativistic limit, such interactions reduce to
\begin{align}\label{eq:W_EM_coupling}
    \mathcal{L}_{int} \sim g^2_{\mathcal{F}\gamma}\Tr[{\bm{\mathcal{F}}}{\bm{\mathcal{F}}}]\,F_{\alpha\beta}\tilde{F}^{\alpha\beta}\,,
\end{align}
where $\bm{\mathcal{F}}=\bW$ and $\bH$ for spin-$1$ and spin-$2$ cases respectively. Here, $F_{\mu\nu}$ is the electromagnetic field strength tensor, and $\tilde{F}_{\mu\nu}$ is its dual. Such couplings may have similar phenomenological implications like the axion photon case \cite{Hertzberg:2018zte,Amin:2021tnq}, but with an important difference due to the polarization state of the soliton. Specifically, $\Tr[{\bm{\mathcal{F}}}{\bm{\mathcal{F}}}]$ is time-independent for  maximally spinning solitons ($\lambda\ne 0)$, but is time-dependent (periodic) for the $\lambda=0$ case and also for  fractionally polarized solitons. Besides emission from a single soliton, a collision between any two non-zero spin solitons may result in radiation that has a specific polarization pattern, depending upon their polarization (for the spin-0 soliton collisions, see for example \cite{Hertzberg:2020dbk,Levkov:2020txo,Amin:2020vja}). 



The above interactions are CP odd. One can have CP even interactions as well, by replacing $F\tilde{F}$ by $F^2$ or other contractions between $F$ and our higher-spin fields. However, one needs to be careful to avoid issues of ghosts and superluminality.\footnote{We would like to thank M.~P.~Hertzberg for bringing these issues to our attention.} Another possibility for the massive spin-$1$ case could be to have kinetic mixing with the usual electromagnetism $\sim F_{\mu\nu}G^{\mu\nu}$.\footnote{We thank Andrew Long for this suggestion.}

\section{Summary and future directions}\label{sec:sum_fd}


Massive, scalar (spin-$0$), vector (spin-$1$) or tensor fields (spin-$2$) can make up all/part of the dark matter, or play a role in the early universe. In this paper, we explored the non-relativistic limit of such fields, and derived polarized ground-state solitons in such fields. We summarize our main results below, and also discuss possible future directions. 
\subsection{Summary}

    \noindent{\bf Non-relativistic Limit}: Starting with the quadratic (free) action for massive spin-$0$,$1$,$2$ fields + leading gravitational interactions, we derived an effective action and equations of motion in the non-relativisitic regime. We arrived at the non-relativistic action  by first using the constraint equations and deriving a quadratic action for physical degrees of freedom in the massive spin-$s$ field. Our non-relativistic system for a field with spin $s$, is a multi-component Schr\"{o}dinger-Poisson (SP) system with $2s+1$ independent d.o.f. 
    \\ \\
    {\bf $\bm{p}$-Solitons}: Using a polarization basis, we enumerate the full space of lowest energy soliton solutions in the SP system for nonzero-spin fields, which we call $p$-solitons. These include extremally polarized solitons which are coherent collections of identically polarized plane waves. The extremally polarized solitons can be used as a basis set to construct fractionally polarized solitons via superposition. All $p$-solitons have a spherically symmetric (universal) energy density profile, although the field configuration is not spherically symmetric. 
    
    The orbital angular momentum is zero for $p$-solitons, but spin angular momentum need not be. In the extremal case, the spin angular momentum is simply the spin multiplicity $\lambda$ times the macroscopically large particle number of the soliton. Explicitly
    \begin{align}
    \bS_{\rm tot}= \lambda \frac{M}{m}\,\,\hat{n}\,,
    \end{align}
where $M\gg m$ is the total mass of the soliton and $m$ is the mass of the dark spin-$s$ field, and the above expression is in units of $\hbar$. The instrinsic spin of our solitons can also have implications for substructure in dark matter today, as well as for the formation of primordial black holes, baryogenesis in the early Universe, etc.
    
For comparison, the average value of a dimensionless (orbital) angular momentum parameter, $\Lambda\equiv J \sqrt{|E|}/G M^{5/2}$, which is used to quantify the angular momentum of dark matter halos in N-body simulations, is $\approx 0.045$ \cite{Vitvitska:2001vw}. For our $p$-solitons, assuming a uniform distribution of the magnitude of the total spin, we have $\langle \Lambda\rangle=\Lambda_{\rm max}/2= 0.12s$. Here, $\Lambda_{\rm max}$ is the value for an extremally polarized soliton and $s$ is the spin of the field. For another comparison, the ratio of the spin angular momentum of our extremal soliton to that of an equal mass black hole is $S_{\rm tot}/J_{\rm bh}=8\pi\lambda\mpl^2/(aMm)$ with $a<1$. This ratio can be larger than unity when $M$ is sufficiently small.\footnote{We used black-hole angular momentum $J_{\rm bh}=a GM^2$.} 
    
    We also compared the energy, particle number and spin of $p$-solitons with other higher energy, but zero total angular momentum solitons with hedgehog configurations of the fields.
    \\ \\
    \noindent {\bf Distinguishability and Implications}: We argued that it is possible to distinguish two interacting $p$-solitons from their scalar counterparts. Limiting ourselves to gravitational interactions in the non-relativistic limit, we discussed how collisions between two higher-spin $p$-solitons cannot be mimicked by two scalar solitons unless the $p$-solitons are identically polarized. Such collisions can in principle be probed by test particles, including photons. We highlighted how non-gravitational couplings to photons can offer further detectability avenues via specific polarization patterns of the outgoing electromagnetic radiation generated by, for example, $p$-soliton collisions.

\subsection{Future Directions}
 \noindent{\bf Formation mechanisms}: Formation mechanisms of solitons in scalar fields have been studied before both with and without self-interactions \cite{Kolb:1993hw,Amin:2011hj,Amin:2019ums}; for higher-spin fields this remains an open question. Coupling the higher-spin fields to a coherently oscillating scalar can lead to efficient production of such fields (for example, see \cite{Agrawal:2018vin,Co:2018lka,Dror:2018pdh}). Higher-spin fields can also be populated via gravitational particle production (albeit with large masses~\cite{Kolb:2020fwh,Alexander:2020gmv}). Solitons could form via self-interactions, or via gravitational clustering in these or in ultra-light fields.  We hope to carry out a numerical investigation of the cosmic history of dark, nonzero-spin fields and formation mechanism of their solitons, which would allow a statistical prediction of their their masses and spins in the early or contemporary universe. For a recent analytic approach for the vector case, see  \cite{Blinov:2021axd}. 
\\ \\
 \noindent{\bf Extension to higher spin fields}: In general, an integer spin-$s$ degree of freedom in embedded in a rank $s$ field (meaning an object carrying $s$ space-time indices). Without mass, the fields are gauge fields (for $s \ge 1$) and depending upon the spin, admit a rich gauge invariance (see~\cite{Bouatta:2004kk,Sorokin:2004ie} and references therein for a classic review). The gauge structure makes sure that there are only two physical $\pm s$ spin multiplicity states. Another equivalent way to construct the Lagrangian density is to add all the possible Lorentz invariant kinetic terms in a way that leads to no ghosts, with only two degrees of freedom surviving (similar to the case of spin-$1$ and spin-$2$). Then, add all possible Lorentz invariant mass terms such that no ghosts, but only the rest $2s-1$ spin multiplicity states now appear. Even though we don't carry out the exercise explicitly for a full Lorentz invariant relativistic theory, we expect that in the non-relativistic and weak field gravity regime, the action is still \eqref{eq:master_action_2}. We have kept most of our expressions in this paper as general as possible, and only in~\ref{sec:grav_bound} have we considered the specific cases of spin-$1$ and spin-$2$ solitons. We leave a detailed study of a fully Lorentz invariant spin-3 and higher fields for future work.
\\ \\
 \noindent{\bf Relativistic corrections}: In this paper we restricted ourselves to the non-relativistic regime, and found $s+1$ {\it distinct}, but degenerate (in energy) soliton configurations for a spin-$s$ bosonic field. The fate of these solutions when relativistic corrections are added, especially for spin-$1$ and spin-$2$, is an important question to consider (for the spin-0 case, see \cite{Salehian:2021khb}). In the non-relativistic limit, we essentially have $2s+1$ spin multiplicity fields with a conserved particle number within each, and thus all of the different $p$-solitons were degenerate and stable. However at sub-leading order, the energy momentum tensor (especially for the cases of spin-$1$ and higher fields) contains non-trivial components due to the spin structure which we do not focus on in this paper. This might lift the degeneracy between the $s+1$ solitons, and affect their stability. In particular, gravitational waves (c.f.~\eqref{eq:interaction_physical}) can be sourced at higher orders, which, depending upon the spin of the $p$-soliton, may differ in their polarization patterns.
 \\ \\

 \noindent{\bf Self-interactions}: It is also possible to include non-gravitational self-interactions within our framework. Attractive self-interactions (even without gravity), are known to support metastable solitons (oscillons) in the scalar case (see, for example, \cite{Copeland:1995fq,Zhang:2020bec}). For the vector case, we pursue the investigation of oscillons held together by self-interactions alone elsewhere~\cite{Zhang:2021xxa}, and find similar $s+1$ stable solitons. An important difference we find is that there is no conserved particle number within {\it each} spin-multiplicity sector. As a result, the extremally polarized solutions are not degenerate in energy, and cannot be superposed to form new additional ones, as we were able to do in the gravitationally supported case. 

In the scalar and vector case, it is possible to find regimes in which either self-interaction, or gravity alone, or a combination of both can be used to support solitons. For the case of massive spin-$2$, self-interactions seem unavoidable and the only ghost free possibility is the full non-linear bi-gravity theory. These self-interactions may become relevant beyond the non-relativistic regime, and must be studied carefully to analyze the fate and lifetimes of our polarized solitons within the full (bi-gravity) theory. We leave a detailed analysis of such questions for future work.\\ \\

\begin{acknowledgments}
 We thank  Siyang Ling, Andrew Long and Rachel Rosen for many useful conversations regarding bi-gravity, as well as Peter Adshead, Mark Hertzberg, Kaloian Lozanov, Zong-Gang Mou and  Hong-Yi Zhang  for helpful discussions and comments on the paper. We are supported by a NASA ATP theory grant NASA-ATP Grant No. 80NSSC20K0518We. 
\end{acknowledgments}
\appendix

\section{}\label{sec:Appendix}

In this appendix, we provide the full non-linear actions for spin-$0$, spin-$1$ and spin-$2$ fields coupled to Einstein gravity. In particular, additional details are provided for some of the issues relevant for a dark, massive spin-$2$ field.

\subsection{Spin-$0$}
In this case, the full nonlinear actions are
\begin{align}\label{eq:0-actions}
    S_{\rm EH} + S_{\rm vis} = \int\mathrm{d}^4x\,\sqrt{-g}\left[-\frac{\mpl^2}{2}R + \mathcal{L}(g,\psi_{\rm SM})\right],\nonumber\\
    S_{\rm dark} = \int\mathrm{d}^4x\sqrt{-g}\left[\frac{1}{2}g^{\mu\nu}\nabla_{\mu}\phi\,\nabla_{\mu}\phi - V(\phi)\right] + ...
\end{align}
where $R$ is the Ricci scalar, and $V(\phi)=(1/2)m^2\phi^2+\lambda_3\phi^3+\hdots$. Expanding the metric around a flat background 
\begin{align}\label{eq:metric_expand1}
    g_{\mu\nu} = \eta_{\mu\nu} + h_{\mu\nu}\,,
\end{align}
and including only leading gravitational interactions, we arrive at the action \eqref{eq:linearized_EH} for $h_{\mu\nu}$ and $\phi$ provided in the main text. There, we also ignored terms beyond quadratic order in $V(\phi)$. These can be included, or ignored, without much difficulty. We can also consider an FLRW background with $\eta_{\mu\nu}$ replaced by and FLRW metric. In this case we must either allow for a background $\phi$ (spatially-independent) or another homogeneous source  for the Friedmann equations. 
\subsection{Spin-$1$}
Our actions $S_{\rm EH}$ and $S_{\rm vis}$ are same as in \eqref{eq:0-actions}, whereas the action for the spin-$1$ field is 
\begin{align}
    S_{\rm dark} &= \int\mathrm{d}^4x\sqrt{-g}\Bigl[-\frac{1}{4}g^{\mu\alpha}g^{\nu\beta}\,G_{\mu\nu}G_{\alpha\beta}\nonumber\\
    &\qquad\qquad\qquad\qquad -V(W_\mu W^\mu)\Bigr] + ...
\end{align}
where $G_{\mu\nu}=\partial_\mu W_\nu-\partial_\nu W_\mu$ and $V(W_\mu W^\mu)=(1/2)m^2W_\mu W^\mu-\lambda_4 (W_\mu W^\mu)^2+\hdots$ could be self interactions arising in the effective low energy theory. Expanding the metric as in \eqref{eq:metric_expand1}, and including only the quadratic part of $V$, yields the action \eqref{eq:L_spin1} for $h_{\mu\nu}$ and $W_{\mu}$ used in the main text. Again, we can also include an FLRW background.
\subsection{Spin-$2$}\label{sec:spin-2_models_A}

For the case of a massive spin-$2$ field, the only ghost-free possibility seems to be the full non-linear bi-metric theory~\cite{deRham:2010kj,Hassan:2011hr,Hassan:2011tf,Hassan:2011zd,Hinterbichler:2011tt,deRham:2014zqa,Schmidt-May:2015vnx} in which there are two metrics, $\mathfrak{g}$ and $\mathfrak{f}$. Composite fluctuations in these metrics give rise to massless and massive spin-$2$ degrees of freedom. Denoting $ S_{\mathfrak{g},\mathfrak{f}}=S_{\rm EH} + S_{\rm dark}$, the bimetric action is
\begin{align}
    S_{\mathfrak{g},f} &= -\frac{\mpl^2}{2(1+\alpha^2)}\int\mathrm{d}^4x\Bigl\{\sqrt{-\mathfrak{g}}\,R(\mathfrak{g}) + \alpha^2\,\sqrt{-\mathfrak{f}}\,R(\mathfrak{f})\nonumber\\
    &\qquad\qquad\qquad\qquad - \frac{2\alpha^2}{1+\alpha^2}m^2\,\sqrt{-\mathfrak{g}}\,\mathcal{V}(X)\Bigr\},
\end{align}
where $\alpha$ is a constant and $X^{\mu}_{\;\;\sigma}X^{\sigma}_{\;\;\nu} = \mathfrak{g}^{\mu\rho}\mathfrak{f}_{\rho\nu}$. The potential $\mathcal{V} = \beta_0 + \beta_1\,X^{\mu}_{\;\;\mu} + \beta_2\,X^{\mu}_{\;\;[\mu}\,X^{\nu}_{\;\;\nu]} + \beta_3\,X^{\mu}_{\;\;[\mu}\,X^{\nu}_{\;\;\nu}\,X^{\alpha}_{\;\;\alpha]} + \beta_4\,X^{\mu}_{\;\;[\mu}\,X^{\nu}_{\;\;\nu}\,X^{\alpha}_{\;\;\alpha}\,X^{\beta}_{\;\;\beta]}$ where $\beta_i$ are constants related by $\beta_1+2\beta_2+\beta_3=1$. We expand both metrics around the flat-spacetime background as\footnote{Note that the metric $\mathfrak{g}$ contains both the massless and massive spin-$2$ excitations (not to be confused with the usual GR metric $g$, appearing in the previous two cases).}
\begin{align}\label{eq:gf_decompose}
    \mathfrak{g}_{\mu\nu} &= \eta_{\mu\nu} + \frac{1}{\mpl}h_{\mu\nu} - \frac{\alpha\sqrt{2}}{\mpl}H_{\mu\nu}\,,\nonumber\\
    \mathfrak{f}_{\mu\nu} &= \eta_{\mu\nu} + \frac{1}{\mpl}h_{\mu\nu} + \frac{\sqrt{2}}{\alpha\,\mpl}H_{\mu\nu}\,,
\end{align}
and obtain quadratic (free) actions for $h$ and $H$ + leading interactions, see eq.~\eqref{eq:bi-gravity_expanded} used in the main text.

We note that in order for $\mathfrak{g}$ and $\mathfrak{f}$ to both have the flat background ansatz (eq.~\eqref{eq:gf_decompose}), we must require the background Einstein equations for both to be consistent with this choice. The Ricci terms ($\sqrt{-\mathfrak{g}}\,R(\mathfrak{g})$ and $\sqrt{-\mathfrak{f}}\,R(\mathfrak{f})$) are consistent. However the potential $\mathcal{V}$ produces cosmological constant terms with $\Lambda_{\mathfrak{g}} = \alpha^2\mpl^2(\beta_0+3\beta_1+3\beta_2+\beta_3)/(1+\alpha^2)$ and $\Lambda_{\mathfrak{f}} = \mpl^2(\beta_1+3\beta_2+3\beta_3+\beta_4)/(1+\alpha^2)$ respectively in the two background equations. We set these to zero to be consistent with our choice~\eqref{eq:gf_decompose}.\footnote{Within bi-gravity (without any matter couplings) with equal backgrounds, the massless $h_{\mu\nu}$ sector alone re-organizes to the usual GR + cosmological constant $\Lambda = \Lambda_\mathfrak{g} = \Lambda_\mathfrak{f}$ \cite{Babichev:2016bxi}.}\\

In general there is also a whole tower of self interactions in the massive spin-$2$ sector. Parametrically, the vertices go like 
\begin{align}\label{eq:self_int_vertices}
    \sim \frac{m^2}{\alpha^n\mpl^{n-2}}H^n \quad\mathrm{and}\quad\sim \frac{m^2\alpha^n}{\mpl^{n-2}}H^n \quad;\quad n \ge 3.\,
\end{align}
In particular, the trilinear interaction vertex ($n=3$ above), is the only coupling that parametrically goes like the leading gravitational coupling ($h_{\mu\nu}\,T^{\mu\nu}(H)/\mpl \sim h\,H^2m^2/\mpl$). All (non-zero) higher order couplings are suppressed by extra factors of $\mpl$. Notice that this coupling cannot lead to on-shell decays of $H$ particle. Moreover in the non-relativistic limit, this vertex contains an oscillatory factor $e^{\pm imt}$ (see \eqref{eq:F_decompose} ahead). As a result, we ignore this 3-point interaction for this paper.
\subsubsection{Coupling to matter}\label{sec:coup_matter}
We also need to worry about the interactions of the massive spin-$2$ field with the SM. Since we have two metrics $\mathfrak{f}$ and $\mathfrak{g}$, and both contain the massive and massless spin-$2$ degrees of freedom (E.q. \eqref{eq:gf_decompose}), other SM fields will couple to both $h$ and $H$ if the metric that appears in the visible action $S_{\rm vis}$ is only $\mathfrak{g}$. This leads to decay channels of type $H\rightarrow{\rm SM} + {\rm SM}$, which puts strong cosmological constraints on $\alpha$~\cite{Babichev:2016bxi}. However, there exists at-least one broader class of metrics in the literature
\begin{align}
    g^{\mathrm{eff}}_{\mu\nu} = a^2\,\mathfrak{g}_{\mu\nu} + 2\,ab\,\mathfrak{g}_{\mu\lambda}X^{\lambda}_{\;\;\nu} + b^2\,\mathfrak{f}_{\mu\nu},
\end{align}
that couple to the usual matter sector, and are ghost free~\cite{Noller:2014sta,Bonifacio:2017nnt}. Explicitly, we have $S_{\rm vis} = \int\mathrm{d}^4x\sqrt{-g_{\mathrm{eff}}}\,\mathcal{L}(g_{\mathrm{eff}},\psi_{\rm SM})$. Expanding $g^{\rm eff}$ using \eqref{eq:gf_decompose} yields 
\begin{align}
    g^{\mathrm{eff}}_{\mu\nu} &= (a+b)^2\left[\eta_{\mu\nu} + \frac{1}{\mpl}h_{\mu\nu} + \frac{\sqrt{2}(b-a\alpha^2)}{\alpha(a+b)\mpl}H_{\mu\nu}\right]\nonumber\\
    &\quad + \mathcal{O}(H^2).
\end{align}
Then with $a=1/(1+\alpha^2)$ and $b={\alpha^2}/{(1+\alpha^2)}$, we can kill the leading order $H_{\mu\nu}$ term in the effective metric (and hence $S_{\rm vis}$). This removes the $H\rightarrow{\rm SM} + {\rm SM}$ channel, alleviating the cosmological bounds on $\alpha$ derived in~\cite{Babichev:2016bxi}. More importantly for our purposes, this allows us to have $\alpha$ values that are neither too large or small and therefore safely neglect the trilinear terms in \eqref{eq:self_int_vertices} in the non-relativistic limit. Furthermore this choice  removes the leading interaction terms between $H_{\mu\nu}$ and other dark sector fields. As a result, one may even consider scenarios in which all different spin dark fields are present in the same theory.
\subsubsection{Cosmological Solutions}
Unlike the spin-$0$ and spin-$1$ cases, including FLRW background comes with additional complications. For example, in an FLRW background there might exist gradient instabilities in the vector perturbations during the radiation dominated era \cite{Comelli:2015pua}. As discussed in that paper, it might be possible to cure these by extending the bi-gravity model.\\

Another thing to consider is the Higuchi bound on the mass of any spin-s field in De-Sitter, $m^2 \ge s(s-1)H^2$ (see~\cite{Lee:2016vti}). For very high scale inflation, this requires the mass of a $s\ge 2$ spin field to be even higher, resulting in ultrasmall solitons. This however is not an issue for sufficiently small scale inflation, with the least conservative bound coming from Big Bang Nucleosynthesis, $H \gtrsim T_{\mathrm{BBN}}/\mpl^2 \sim 10^{-16}$ eV.\\

Furthermore, the cut-off of the bi-gravity theory  $\Lambda_3\sim (m^2\mpl)^{1/3}$ (obtained for example, by considering unitarity bounds from $2\rightarrow 2$ scattering of various helicity modes of the field) can be quite small when $m$ is sufficiently small. While this may not be an issue in the non-relativistic limit where $k\ll m$, a consistent, calculable set-up at higher energies which become accessible in the early universe, might require further work. 


\bibliography{references}
\end{document}